\documentclass[letterpaper,twocolumn,10pt]{article}
\usepackage{usenix}
\usepackage{times}        

\usepackage[subtle,tracking=normal,margins=normal,leading=normal,title=tight]{savetrees-clean}

\usepackage{booktabs}       
\usepackage{comment}        
\usepackage{enumitem}       
\usepackage{epsfig}         
\usepackage{etoolbox}       
\usepackage[hidelinks]{hyperref}
\usepackage{paralist}       
\usepackage{subcaption}     
\usepackage{titling}        
\usepackage[normalem]{ulem}   
\usepackage{upgreek}        
\usepackage{url}            
\usepackage{xcolor}         

\newtoggle{showmarks}

\definecolor{darkgreen}{rgb}{0,0.7,0}
\definecolor{darkred}{rgb}{0.7,0,0}
\definecolor{orange}{rgb}{1,0.4,0}

\iftoggle{showmarks}{
\newcommand\red[1]{\textcolor{red}{#1}}
\newcommand\redstrike[1]{\red{\sout{#1}}}
\newcommand\green[1]{\textcolor{darkgreen}{#1}}
\newcommand\greenstrike[1]{\green{\sout{#1}}}
\newcommand\greenstrikealt[1]{\green{\sout{#1}}}
\newcommand\orange[1]{\textcolor{orange}{#1}}
\newcommand\orangestrike[1]{\orange{\sout{#1}}}
\newcommand\blue[1]{\textcolor{blue}{#1}}
\newcommand\bluestrike[1]{\blue{\sout{#1}}}
\newcommand\MA[1]{\textcolor{red}{[\bf MA: #1]}}
\newcommand\todo[1]{\textcolor{blue}{(TODO: #1)}}
}{
\newcommand\red[1]{#1}
\newcommand\redstrike[1]{\unskip}
\newcommand\green[1]{#1}
\newcommand\greenstrike[1]{\unskip}
\newcommand\greenstrikealt[1]{}
\newcommand\orange[1]{#1}
\newcommand\orangestrike[1]{\unskip}
\newcommand\blue[1]{\unskip}
\newcommand\bluestrike[1]{\unskip}
\newcommand\MA[1]{\unskip}
\newcommand\todo[1]{\unskip}
}

\hypersetup{colorlinks=true,
        linkcolor=darkred,
        citecolor=darkgreen}

\usepackage{caption}
\usepackage{titlesec}
\titlespacing*{\section}{0pt}{7pt plus 3pt minus 3pt}{3pt plus 3pt minus 2pt}
\titlespacing*{\subsection}{0pt}{4pt plus 3pt minus 2pt}{1pt plus 3pt minus 1pt}
\titlespacing*{\subsubsection}{0pt}{4pt plus 3pt minus 2pt}{0pt plus 3pt minus 1pt}
\titleformat{\section}{\large\bfseries}{\thesection}{1em}{}
\titleformat{\subsection}{\normalsize\bfseries}{\thesubsection}{1em}{}

\renewcommand\smallskip{\vspace{2pt}}

\newcommand\figSqueeze{0.0mm}

\newcommand*{\helvetica}{\fontfamily{phv}\selectfont}

\graphicspath{{./}{figures/}}
\begin{document}

\setlength{\droptitle}{-0.9in}

\title{\Large\bf
Homa: A Receiver-Driven Low-Latency\\
Transport Protocol Using Network Priorities (Complete Version)\vspace{-0.2in}}

\author{
Behnam Montazeri, Yilong Li, Mohammad Alizadeh\textsuperscript{\dag},
and John Ousterhout
\\Stanford University, \textsuperscript{\dag}MIT\vspace{0.2in}
\\\normalsize{\bf
This paper is an extended version of the paper on Homa that was published in ACM
SIGCOMM 2018.}
\\\normalsize{\bf
Material had to be removed from Sections 5.1 and 5.2 to meet the
SIGCOMM page restrictions; this}
\\\normalsize{\bf
version restores the missing material.}
}
\vspace{0.5in}

\date{\vspace{-0.2in}}

\maketitle

\begin{abstract}
Homa is a new transport protocol for datacenter networks. It provides
exceptionally low latency, especially for workloads with a high
volume of very short messages, and it also supports large messages and
high network utilization. Homa uses in-network
priority queues to ensure low latency for short messages; priority
allocation is managed dynamically by each receiver and integrated
with a receiver-driven flow control mechanism. Homa also uses
controlled overcommitment of receiver downlinks to ensure efficient
bandwidth utilization at high load.
Our implementation of Homa delivers 99th percentile round-trip times
less than 15 $\upmu$s for short messages on a 10 Gbps network
running at 80\% load. These latencies are almost 100x lower than
the best published measurements of an implementation. In simulations,
Homa's latency is roughly equal to pFabric and significantly better than
pHost, PIAS,
and NDP for almost all message sizes and workloads. Homa can also sustain
higher network loads than pFabric, pHost, or PIAS.
\end{abstract}

\section{Introduction}
\label{sec:intro}

The rise of datacenter computing over the last decade has created
new operating conditions for network transport protocols.
Modern datacenter networking hardware offers the potential for very
low latency communication. Round-trip times of 5 $\upmu$s or less are
now possible for short messages, and a variety of applications have
arisen that can take advantage of this
latency~\cite{memcached,redis-www,ramcloud-tocs}.
In addition, many datacenter applications use request-response
protocols that are dominated by very
short messages (a few hundred bytes or less). Existing transport
protocols are ill-suited to these conditions, so the latency
they provide for short messages is far higher
than the hardware potential, particularly under high network loads.

Recent years have seen numerous proposals for better transport
protocols, including improvements to TCP \cite{dctcp,hull,d2tcp}
and a variety of new protocols \cite{d3,pdq,pfabric,fastpass,qjump,pias,ndp}.
However, none of these designs considers today's small message
sizes; they are based on heavy-tailed workloads where 100 Kbyte
messages are considered ``small,'' and latencies are often
measured in milliseconds, not microseconds.
As a result, there is still no practical solution that
provides near-hardware latencies for short messages under high
network loads. For example, we know of no existing
implementation with tail
latencies of 100 $\upmu$s or less at high network
load (within 20x of the hardware potential).

Homa is a new transport protocol designed for
small messages in low-latency datacenter environments.
Our implementation of Homa achieves 99th percentile round trip latencies less
than 15 $\upmu$s for small messages at 80\% network load with
10 Gbps link speeds, and it does this even in the presence of
competing large messages. Across a wide range of message sizes
and workloads, Homa achieves 99th percentile latencies at
80\% network load that are within a factor of 2--3.5x of the
minimum possible latency on an unloaded network. Although Homa
favors small messages, it also improves the performance of large
messages in comparison to TCP-like approaches based on fair
sharing.

Homa uses two innovations to achieve its high performance. The
first is its aggressive use of the priority queues provided by modern
network switches. In order to make the most of the limited
number of priority queues, Homa assigns priorities dynamically
on receivers, and it integrates the priorities with a receiver-driven
flow control mechanism like that of pHost~\cite{phost} and NDP~\cite{ndp}.
Homa's priority mechanism improves tail latency by 2--16x compared
to previous receiver-driven approaches.
In comparison to sender-driven priority mechanisms such as PIAS~\cite{pias},
Homa provides a better approximation to SRPT (shortest remaining
processing time first); this reduces tail latency by 0--3x over PIAS.

Homa's second contribution is its use of
\emph{controlled overcommitment}, where a receiver allows a few
senders to transmit simultaneously. Slightly overcommitting
receiver downlinks in this way allows Homa to use network bandwidth
efficiently: Homa can sustain network loads 2--33\% higher than
pFabric~\cite{pfabric}, PIAS, pHost, and NDP.
Homa limits the overcommitment and integrates it with the priority
mechanism to prevent queuing of short messages.

Homa has several other unusual features that contribute to
its high performance.
It uses a message-based architecture rather than a streaming
approach; this eliminates
head-of-line blocking at senders and reduces tail latency by 100x
over streaming transports such as TCP. Homa
is connectionless, which
reduces connection state in large-scale applications.
It has no explicit acknowledgments,
which reduces overheads for small messages, and it implements
at-least-once semantics rather than at-most-once.

\section{Motivation and Key Ideas}
\label{sec:motivation}

The primary goal of Homa is to provide the lowest possible latency for short messages at high network load using current networking hardware. We focus on tail message latency (99th percentile), as it is the most important metric for datacenter applications~\cite{dctcp, detail}. A large body of work has focused on low latency datacenter transport in recent years. However, as our results will show, existing designs are sub-optimal for tail latency at high network load, particularly in networks with raw hardware latency in the single-digit microseconds~\cite{shanley2003infiniband, dpdk, dcqcn, timely}. In this section, we discuss the challenges that arise in such networks and we derive Homa's key design features.


\subsection{Motivation: Tiny latency for tiny messages}

State-of-the-art cut-through switches have latencies of at most a few hundred nanoseconds~\cite{tomahawk}. Low latency network interface cards and software stacks (e.g., DPDK~\cite{dpdk}) have also become common in the last few years. These advances have made it possible to achieve one-way latencies of a few microseconds in the absence of queuing, even across a large network with thousands of servers (e.g., a 3-level fat-tree network). 

\begin{figure}
\begin{center}
{\footnotesize
\centering
\begin{tabular}{ c p{0.3\textwidth}}
W1 & Accesses to a collection of memcached servers at
Facebook, as approximated by the statistical model of the ETC workload
in Section 5 and Table 5 of ~\cite{atikoglu-kvs}.
\\
W2 & Search application at Google~\cite{googleStats}.
\\
W3 & Aggregated workload from all applications running in a
Google datacenter~\cite{googleStats}.
\\
W4 & Hadoop cluster at Facebook~\cite{hadoopStats}.
\\
W5 & Web search workload used for DCTCP~\cite{dctcp}.
\\
\end{tabular}
}
\vspace{1.0ex}
\includegraphics[scale=0.4]{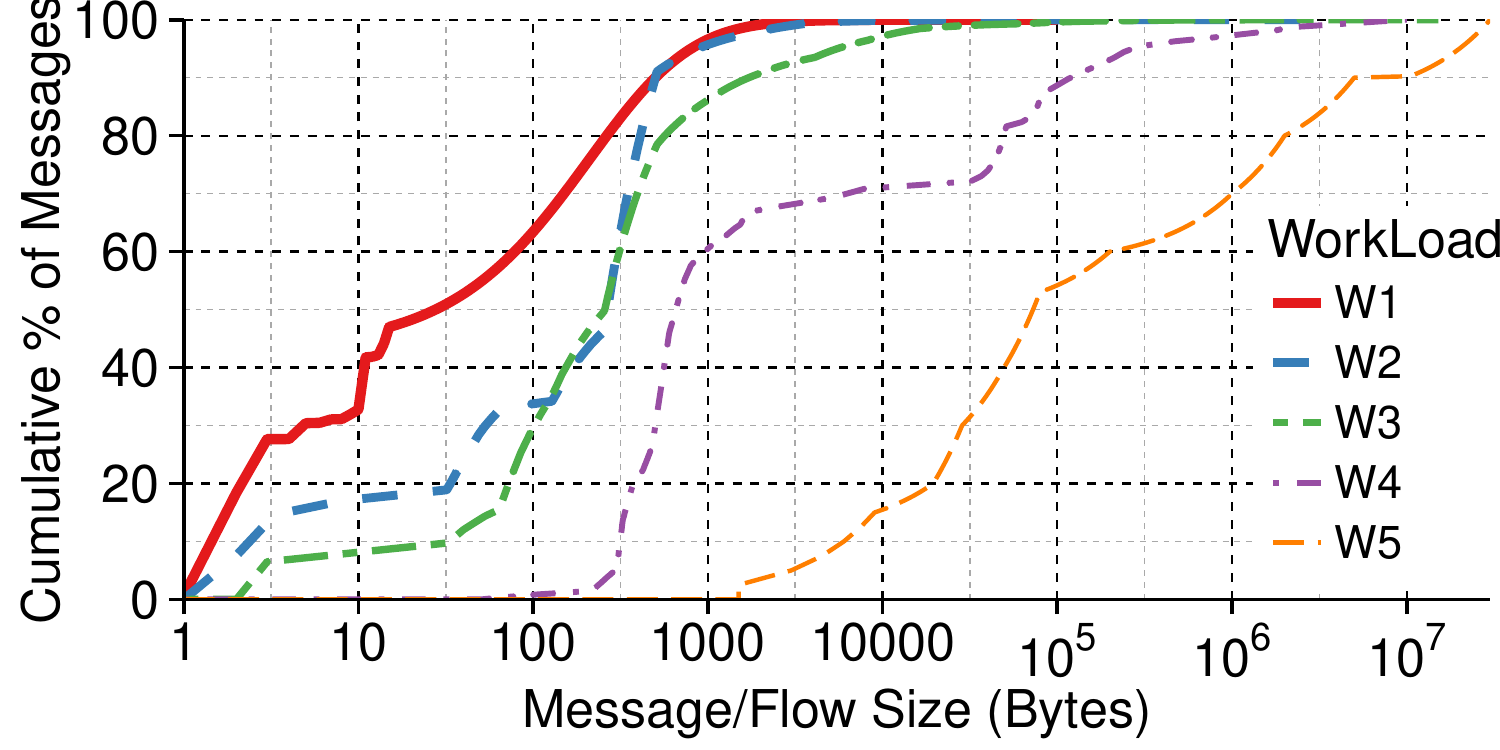}
\vspace{1.0ex}
\includegraphics[scale=0.4]{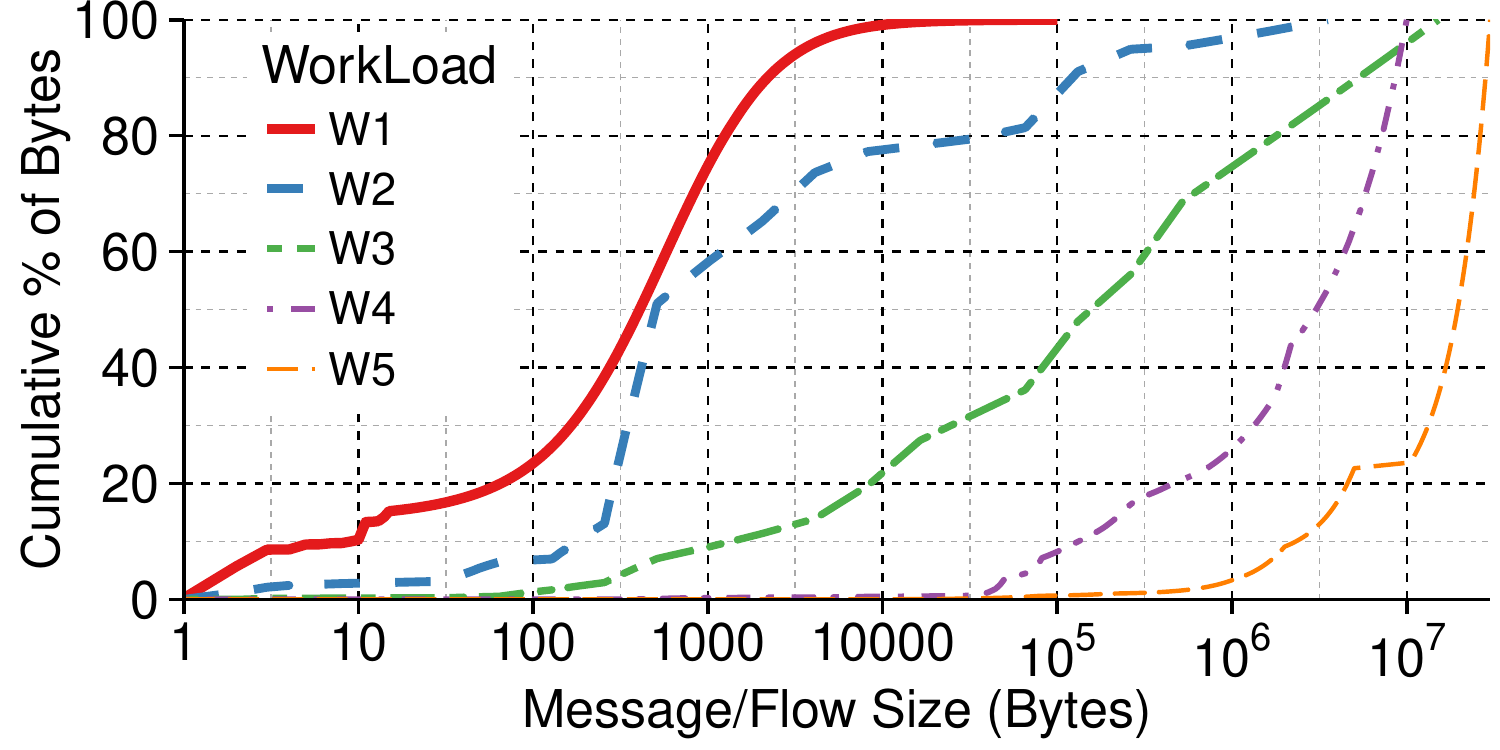}
\caption{The workloads used to design and evaluate Homa. Workloads
W1--W3 were measured from application-level logs of message sizes;
message sizes for W4 and W5 were estimated from packet traces. The
upper graph shows the cumulative distribution of message sizes
weighted by number of messages, and the lower graph is weighted by
bytes. The workloads are ordered by average message size: W1
is the smallest, and W5 is most heavy-tailed.}
\label{fig:workloads}
\vspace{\figSqueeze{}}
\end{center}
\end{figure}

Meanwhile, many datacenter applications rely on request-response protocols with tiny messages of {\em a few hundred bytes or less}. In typical remote procedure call (RPC) use cases, it is almost always the case that either the request or the response is tiny, since data usually flows in only one direction. The data itself is often very short as well. Figure~\ref{fig:workloads} shows a collection of workloads that we used to design and evaluate Homa, most of which were measured from datacenter applications at Google and Facebook. In three of these workloads, more than 85\% of messages were less than 1000 bytes. In the most extreme case (W1), more than 70\% of all network traffic, measured in bytes, was in messages less than 1000 bytes.

To our knowledge, almost all prior work has focused on workloads with very large messages. For example, in the Web Search workload used to evaluate DCTCP~\cite{dctcp} and pFabric~\cite{pfabric} (W5 in Figure~\ref{fig:workloads}), messages longer than 1 Mbyte account for 95\% of transmitted bytes, and any message shorter than 100~Kbytes was considered ``short.'' Most subsequent work has used the same workloads. To obtain these workloads, message sizes were estimated from packet captures based on inactivity of TCP connections beyond a threshold (e.g., 50~ms). Unfortunately, this approach overestimates message sizes, since a TCP connection can contain many closely-spaced messages. In Figure~\ref{fig:workloads}, workloads W1--W3 were measured explicitly in terms of application-level messages, and they show much smaller sizes than workloads W4 and W5, which were extracted from packet captures.

Unfortunately, existing datacenter transport designs cannot achieve the lowest possible latency for tiny messages at high network load. We explore the design space in the next section, but consider, for example, designs that do not take advantage of in-network priorities (e.g., HULL~\cite{hull}, PDQ~\cite{pdq}, NDP~\cite{ndp}). These designs attempt to limit queue buildup, but none of them can eliminate queuing altogether. The state-of-the-art approach, NDP~\cite{ndp}, strictly limits queues to 8 packets, equivalent to roughly 10~$\mu$s of latency at 10~Gbps. While this queuing latency has negligible impact in a network with moderate latency (e.g., RTTs greater than 50~$\upmu$s) or for moderately-sized messages (e.g., 100 KBytes), it increases by 5x the completion time of a 200-byte message in a network with 5~$\mu$s RTT.

\subsection{The Design Space}
We now present a walk through the design space of low latency datacenter transport protocols. We derive Homa's four key design principles: (i) transmitting short messages blindly, (ii) using in-network priorities, (iii) allocating priorities dynamically at receivers in conjunction with receiver-driven rate control, and (iv) controlled overcommitment of receiver downlinks. While some past designs use the first two of these techniques, we show that combining all four techniques is crucial to deliver the lowest levels of latency at high network load.

We focus on {\em message} latency (not packet latency) since it reflects application performance. A message is a block of bytes of any length transmitted from a single sender to a single receiver. The sender must specify the size of a message when presenting its first byte to the transport,
and the receiver cannot act on a message until it has been received
in its entirety. Knowledge of message sizes is particularly valuable
because it allows transports to prioritize shorter messages.

The key challenge in delivering short messages with low latency is to
eliminate queuing delays.
Similar to prior work, we assume that bandwidth in the network
core is sufficient to accommodate the offered load, and that the network
supports efficient load-balancing~\cite{packetSpraying, presto, conga},
so that packets are distributed evenly across the available paths (we
assume simple randomized per-packet spraying in our design). As a result,
queueing will occur primarily in the downlinks from top-of-rack switches
(TORs) to machines. This happens when multiple senders
transmit simultaneously to the same receiver. The worst-case
scenario is \emph{incast}, where an application initiates RPCs
to many servers concurrently and the responses all arrive
at the same time.


\smallskip
\noindent{\bf There is no time to schedule every packet.}
An ideal scheme might attempt to schedule every packet at a central arbiter, as in Fastpass~\cite{fastpass}. Such an arbiter could take into account
all the messages and make a global scheduling decision about which packet to transmit from each sender and when to transmit it. The arbiter could
in theory avoid queues in the network altogether. However, this
approach triples the latency for
short messages: a tiny, single-packet message takes at least 1.5
RTTs if it needs to wait for a scheduling decision, whereas it could
finish within 0.5 RTT if transmitted immediately. Receiver-based
scheduling mechanisms such as ExpressPass~\cite{express} suffer the
same penalty.

In order to achieve the lowest possible latency, short messages must be
transmitted {\em blindly}, without considering potential congestion.
In general, a sender must transmit enough bytes blindly
to cover the round-trip time to the receiver (including software
overheads on both ends); during this time the receiver can return
explicit scheduling information to control future
transmissions, without introducing additional delays. We refer to
this amount of data as \emph{RTTbytes}; it is about 10~KB in our implementation of Homa for 10~Gbps networks.

\smallskip
\noindent{\bf Buffering is a necessary evil.}
Blind transmissions mean that buffering can occur when multiple
senders transmit to the same receiver. No protocol can achieve minimum
latency without incurring some buffering. But, ironically, when
buffering occurs, it will increase latency. Many previous designs have attempted to reduce buffering, e.g., with carefully-engineered rate control schemes~\cite{dctcp, dcqcn, timely}, reserving bandwidth headroom~\cite{hull}, or even strictly limiting the buffer size to a small value~\cite{ndp}. However, none of these approaches can completely eliminate the latency penalty of buffering. 


\smallskip
\noindent{\bf In-network priorities are a must.}
Given the inevitability of buffering, the only way to achieve the
lowest possible latency is to use in-network priorities.
Each output port in a modern switch supports a small number of priority levels
(typically 8), with one queue for each priority. Each
incoming packet indicates which queue to use for that packet, and
output ports service higher priority queues before lower priority ones.
The key to low latency is assigning packet priorities so that short
messages bypass queued packets for longer messages.

This observation is not new; starting with pFabric~\cite{pfabric},
several schemes have shown that switch-based priorities can be
used to improve message latency~\cite{qjump, phost, pias, karuna}. These schemes use priorities to implement various message-size-based scheduling policies. The most common of these policies is SRPT (shortest remaining processing time first), which prioritizes packets from messages with the fewest bytes remaining to transmit. SRPT provides near-optimal average message latency, and as shown in prior work~\cite{pdq, pfabric}, it also provides very good tail latency for short messages. Homa implements an approximation of SRPT (though the design can support other policies as well).

Unfortunately, in practice, no existing scheme can deliver the near-optimal latency of SRPT at high network load.  pFabric approximates SRPT accurately, but it requires too many priority levels to implement with today's switches. PIAS~\cite{pias} works with a limited number of priorities, but it assigns priorities on senders, which limits its ability to approximate SRPT (see below). In addition, it works without message sizes, so it uses a ``multi-level queue'' scheduling policy. As a result, PIAS has high tail latency both for short messages and long ones. QJUMP~\cite{qjump} requires priorities to be allocated manually on a per-application basis, which is too inflexible to produce optimal latencies. 

\smallskip
\noindent{\bf Making best use of limited priorities requires receiver control.} To produce the best approximation of SRPT with only a small number of priority levels, the priorities should be determined by the receiver. Except for blind transmissions, the receiver knows the exact set of messages vying for bandwidth on its downlink from the TOR switch. As a result, the receiver can best decide which priority to use for each incoming packet. In addition, the receiver can amplify the effectiveness of the priorities by integrating them with a packet scheduling mechanism.

pHost~\cite{phost}, the closest prior scheme to Homa, is an example of using a receiver-driven approach to approximate SRPT. Its primary mechanism is packet scheduling: senders transmit the first RTTbytes of each message blindly, but packets after that are transmitted only in response to explicit {\em grants} from the receiver. Receivers schedule the grants to implement SRPT while controlling the influx of packets to match the downlink speed.

However, pHost makes only limited use of priorities: it statically assigns one high priority for all blind transmissions and one lower priority for all scheduled packets. This impacts its ability to approximate SRPT in two ways. First, it bundles all blind transmissions into a single priority. While this is reasonable for workloads where most bytes are from large messages (W4-W5 in Figure~\ref{fig:workloads}), it is problematic for workloads where a large fraction of bytes are transmitted blindly (W1-W3). Second, for messages longer than RTTbytes, pHost cannot preempt a larger message immediately for a shorter one. Once again, the root of the problem is that pHost bundles all such messages into a single priority, which results in queueing delays. We will show in \S\ref{sec:priorities} that this creates {\em preemption lag}, which hurts latency, particularly for medium-sized messages that last a few RTTs.

\smallskip
\noindent{\bf Receivers must allocate priorities dynamically.}
Homa addresses pHost's limitations by dynamically allocating multiple priorities at the receivers.  Each receiver allocates priorities for its own downlink using two mechanisms. For messages larger than RTTbytes,  the receiver communicates a priority for each packet to its sender dynamically based on the exact set of inbound messages. This eliminates almost all preemption lag. For short messages sent blindly, the sender cannot know about other messages inbound for the receiver. Even so, the receiver can provide guidance in advance to senders based on its recent workload. Our experiments show that dynamic priority management reduces tail latency considerably in comparison to  static priority allocation schemes such as those in pHost or PIAS.

\smallskip
\noindent{\bf Receivers must overcommit their downlink in a controlled manner.} Scheduling packet transmissions with grants from receivers reduces buffer occupancy, but it introduces a new challenge: a receiver may send grants to a sender that does  not transmit to it in a timely manner. This problem occurs, for instance, when a sender has messages for multiple receivers; if more than one receiver decides to send it grants, the sender cannot transmit packets to all such receivers at full speed. This wastes bandwidth at the receiver downlinks and can significantly hurt performance at high network load. For example, we find that the maximum load that pHost can support ranges between 58\% and 73\% depending on the workload, despite using a timeout mechanism to mitigate the impact of unresponsive senders (\S\ref{sec:simulations}). NDP~\cite{ndp} also schedules incoming packets to avoid buffer buildup, and it suffers from a similar problem.

To address this challenge, Homa's receivers intentionally \emph{overcommit} their downlinks by granting simultaneously to a small number of senders; this results in controlled packet queuing at the receiver's TOR but is crucial to achieve high network utilization and the best message latency at high load (\S\ref{sec:overcommit}).

\smallskip
\noindent{\bf Senders need SRPT also.} Queues
can build up at senders as well as receivers, and this can result in
long delays for short messages. For example, most existing protocols
implement byte streams, and an application will
typically use a single stream for each destination. However, this
can result in head-of-line-blocking, where a short message for a
given destination is queued in the byte stream behind a long message for
the same destination. \S\ref{sec:implementationEval} will show
that this increases tail latency by 100x for short messages. FIFO packet
queues in the NIC can also result in high tail latency for short messages,
even if messages are transmitted on different streams. For low tail
latency, senders must ensure that short outgoing messages are
not delayed by long ones.

\begin{figure}
\begin{center}
\includegraphics[scale=0.45]{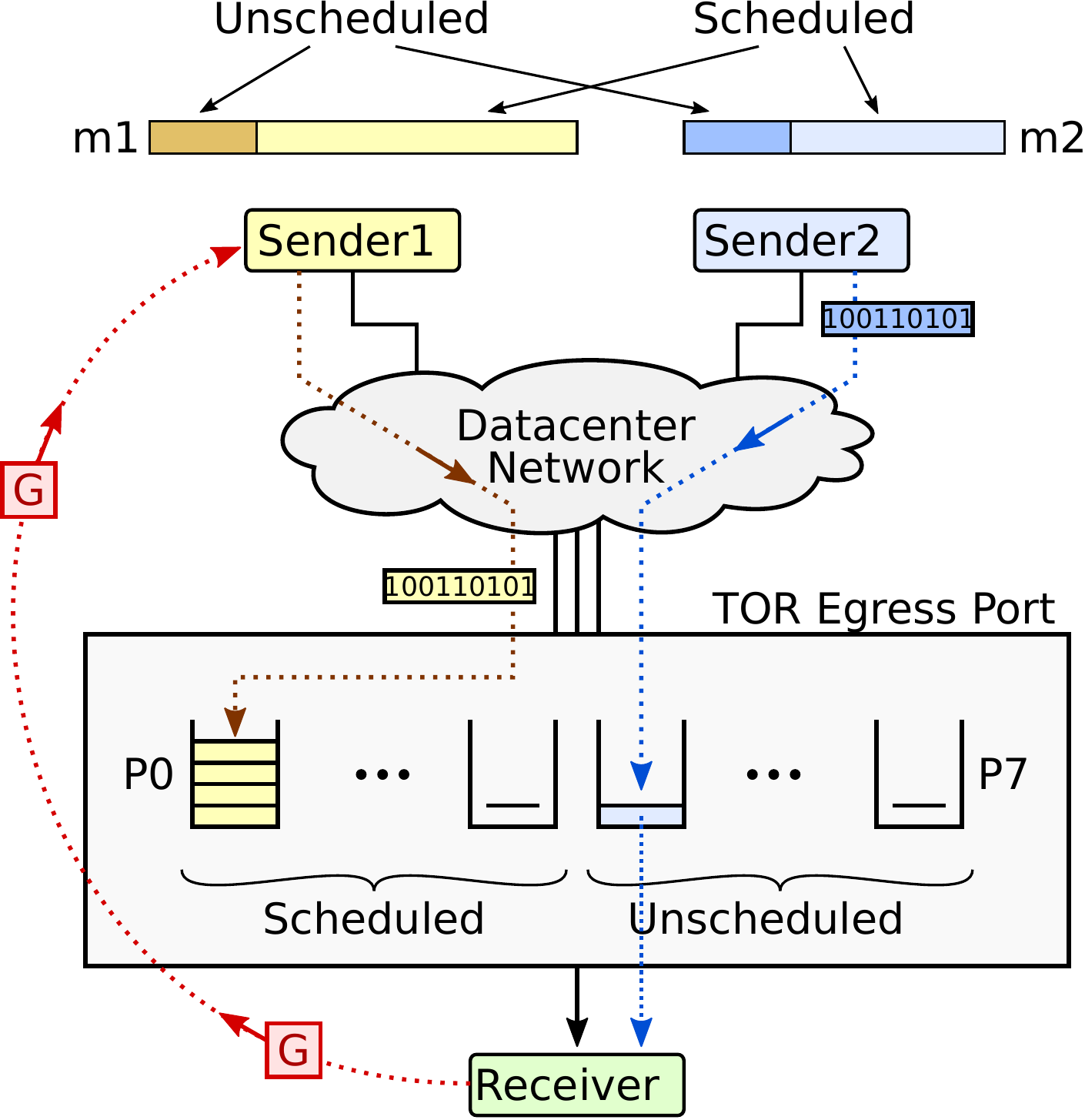}
\caption{Overview of the Homa protocol. Sender1 is transmitting
scheduled packets of message \emph{m1},
while Sender2 is transmitting unscheduled packets of \emph{m2}.
}
\label{fig:overview}
\vspace{\figSqueeze{}}
\end{center}
\end{figure}

\smallskip
\noindent{\bf Putting it all together.} Figure~\ref{fig:overview} shows an overview of the Homa protocol. Homa divides messages into two parts: an initial \emph{unscheduled} portion followed by a \emph{scheduled} portion.
The sender transmits the unscheduled packets (RTTbytes of data) immediately, but it does not transmit any scheduled packets until instructed by the receiver. The arrival of an unscheduled packet makes the receiver aware of the message; the receiver then requests the transmission of scheduled packets by sending one \emph{grant} packet for each scheduled packet. Homa's receivers dynamically set priorities for scheduled packets and periodically notify senders of a set of thresholds for setting priorities for unscheduled packets. Finally, the receivers implement controlled overcommitment to sustain high utilization in the presence of unresponsive senders. The net effect is an accurate approximation of the SRPT scheduling policy using a small number of priority queues. We will show that this yields excellent performance across a broad range of workloads and traffic conditions.

\section{Homa Design}
\label{sec:design}

This section describes the Homa protocol in detail. In addition
to describing how Homa implements the key ideas from the previous section,
this section also discusses several other aspects of the protocol
that are less essential for performance but result in a complete
and practical substrate for datacenter RPC. Homa contains several
unusual features: it is receiver-driven; it is message-oriented,
rather than stream-oriented; it is connectionless; it uses no explicit
acknowledgments; and it implements at-least-once semantics, rather than
the more traditional at-most-once semantics. Homa uses four packet
types, which are summarized in Figure~\ref{fig:packetTypes}.

\begin{figure}
\footnotesize{
\begin{description}[style=multiline,leftmargin=1.5cm]
\item[DATA] Sent from sender to receiver. Contains a range
of bytes within a message, defined by an offset and a length.
Also indicates total message length.
\item[GRANT] Sent from receiver to sender.
Indicates that the sender may now transmit all bytes in the message
up to a given offset, and specifies the
priority level to use.
\item[RESEND] Sent from receiver to sender. Indicates that sender
should re-transmit a given range of bytes within a message.
\item[BUSY] Sent from sender to receiver.
Indicates that a response to RESEND will be delayed (the
sender is busy transmitting higher priority messages, or an RPC
operation is still being executed); used to prevent timeouts.
\end{description}
}
\begin{center}
\caption{
The packet types used by Homa. All packet types except DATA are
sent at highest priority; the priorities for DATA packets are specified
by the receiver as discussed in \S\ref{sec:priorities}.
}
\label{fig:packetTypes}
\vspace{\figSqueeze{}}
\end{center}
\end{figure}

\subsection{RPCs, not connections}
\label{subsec:noConnections}

Homa is connectionless. It implements the basic data transport for
RPCs, each of which consists
of a \emph{request message} from a \emph{client} to a \emph{server}
and its corresponding \emph{response message}. Each
RPC is identified by a globally unique \emph{RPCid}
generated by the client. The RPCid is included in all packets associated
with the RPC.  A client may have any number of outstanding RPCs at a
time, to any number of servers; concurrent RPCs to the same server may
complete in any order.

Independent delivery of messages is essential for low tail latency.
The streaming approach used by TCP results in head-of-line-blocking,
where a short message is queued behind a long message for the same
destination. \S\ref{sec:implementationEval} will show
that this increases tail latency by 100x for short messages. Many
recent proposals, such as DCTCP, pFabric,
and PIAS, assume dozens of connections between each source-target pair,
so that each messsage has a dedicated connection. However, this approach
results in an explosion of connection state. Even a single connection
for each application-server pair is problematic for large-scale
applications (\cite{scalingMemcached} \S3.1, \cite{farm-nsdi} \S3.1), so it
is probably not realistic to use multiple connections.

No setup phase or connection is required before a client initiates an
RPC to a server, and neither the client nor the server retains any state
about an RPC once the client has received the result. In datacenter
applications, servers can have large numbers of
clients; for example, servers in Google datacenters commonly
have several hundred thousand open connections \cite{feldermanConnections}.
Homa's connectionless
approach means that the state kept on a server is determined by the
number of active RPCs, not the total number of clients.

Homa requires a response for each RPC request because this is the
common case in datacenter applications and it allows the response to
serve as an acknowledgment for the request. This reduces the number
of packets required (in the simplest case, there is only a single
request packet and a single response packet). One-way messages
can be simulated by having the server application return an empty
response immediately upon receipt of the request.

Homa handles request and response messages in nearly identical fashion,
so we don't distinguish between requests and responses in most of the
discussion below.

Although we designed Homa for newer datacenter applications
where RPC is a natural fit, we believe that traditional applications
could be supported by implementing a socket-like byte stream interface
above Homa. We leave this for future work.

\subsection{Basic sender behavior}
\label{sec:senderBehavior}

When a message arrives at the sender's transport module, Homa divides
the message into two parts: an initial \emph{unscheduled} portion
(the first RTTbytes bytes), followed by a \emph{scheduled} portion.
The sender transmits the unscheduled bytes immediately, using one or
more DATA packets. The scheduled bytes are not transmitted until
requested explicitly by the receiver using GRANT packets.
Each DATA packet has a priority, which is determined by the
receiver as described in \S\ref{sec:priorities}.
 
The sender implements SRPT for its outgoing packets: if DATA packets
from several messages are ready for transmission
at the same time, packets for the message with the fewest remaining
bytes are sent first. The sender does not consider the priorities in
the DATA packets when scheduling its packet transmissions (the priorities
in DATA packets are intended for the final downlinks to the receivers).
Control packets such as GRANTs and RESENDs are always given priority over
DATA packets.

\subsection{Flow control}

Flow control in Homa is implemented on the receiver side
by scheduling incoming packets on a
packet-by-packet basis, like pHost and NDP. Under
most conditions, whenever a
DATA packet arrives at the receiver, the receiver sends a
GRANT packet back to the sender. The grant invites the sender
to transmit all bytes in the message up to a given offset,
and the offset is chosen so that there are always RTTbytes
of data in the message that have been granted but not yet
received. Assuming timely delivery of grants back to the sender
and no competition from other messages,
messages can be transmitted from start to finish at line rate
with no delays.

If multiple messages arrive at a receiver simultaneously, their
DATA packets will interleave as determined by their priorities.
If the DATA packets of a message are delayed, then GRANTs for that
message will also be delayed, so there will never be more than
RTTbytes of granted-but-not-received data for a message. This
means that each incoming message can occupy at most RTTbytes of
buffer space in the receiver's TOR.

If there are multiple incoming messages, the receiver may stop
sending grants to some of them, as part of the overcommitment
limits described in \S\ref{sec:overcommit}. Once a grant
has been sent for the last bytes of a message, data packets for that
message may result in grants to other messages for which
grants had previously been stopped.

The DATA packets for a message can arrive in any order; the
receiver collates them using the offsets in each packet.
This allows Homa to use per-packet multi-path routing
in order to minimize congestion in the network core.

\subsection{Packet priorities}
\label{sec:priorities}

The most novel feature in Homa, and the key to its performance,
is its use of priorities.
Each receiver determines the priorities for all of its incoming
DATA packets in order to approximate the SRPT policy.
It uses different mechanisms for unscheduled and scheduled packets.
For unscheduled packets, the receiver allocates priorities in advance.
It uses recent traffic patterns to choose priority allocations, and it
disseminates that information to senders by piggybacking it on
other packets. Each sender retains the most recent allocations for each
receiver (a few dozen bytes per receiver) and uses that
information when transmitting unscheduled packets. If the
receiver's incoming traffic changes, it disseminates new priority
allocations the next time it communicates with each sender.

\begin{figure}
\begin{center}
\includegraphics[scale=0.32]{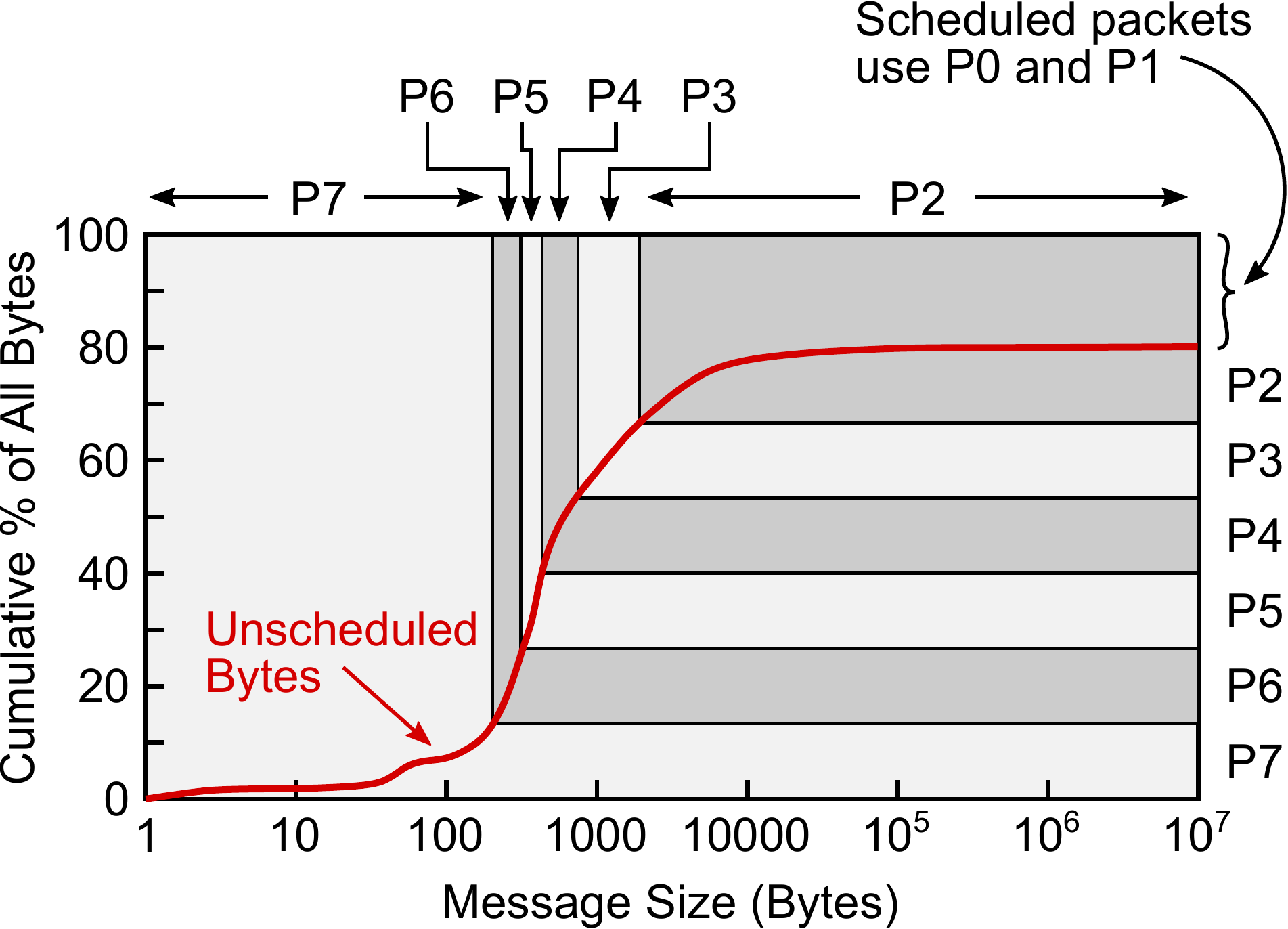}
\caption{Homa receivers allocate unscheduled priorities based on traffic
patterns. This figure shows the CDF of unscheduled bytes across
messages of different sizes for workload W2; 100\% on the y-axis
corresponds to \emph{all} network traffic, both scheduled and unscheduled.
About 80\% of all bytes are unscheduled; Homa allocates a corresponding
fraction of priority levels (6 out of 8) for unscheduled packets.
The CDF is then used to determine the range of message sizes for each
priority level so that traffic is evenly distributed among them. For example,
P7 (the highest priority level) will be used for unscheduled bytes for
messages of length 1--280 bytes.
}
\label{fig:priorityAllocation}
\vspace{\figSqueeze{}}
\end{center}
\end{figure}

Homa allocates priorities for unscheduled packets so that each
priority level is used for about the same number of bytes. Each
receiver records statistics about the sizes of its incoming messages
and uses the message size distribution to compute priority levels
as illustrated in Figure~\ref{fig:priorityAllocation}. The
receiver first computes the fraction of
all incoming bytes that are unscheduled (about 80\% in
Figure~\ref{fig:priorityAllocation}). It allocates this fraction
of the available priorities (the highest ones) for unscheduled
packets, and reserves the remaining (lower) priority levels for scheduled
packets. The receiver then chooses cutoffs between the
unscheduled priorities so that each priority
level is used for an equal number of unscheduled bytes and shorter
messages use higher priorities.

\begin{figure}
\begin{center}
\includegraphics[scale=0.45]{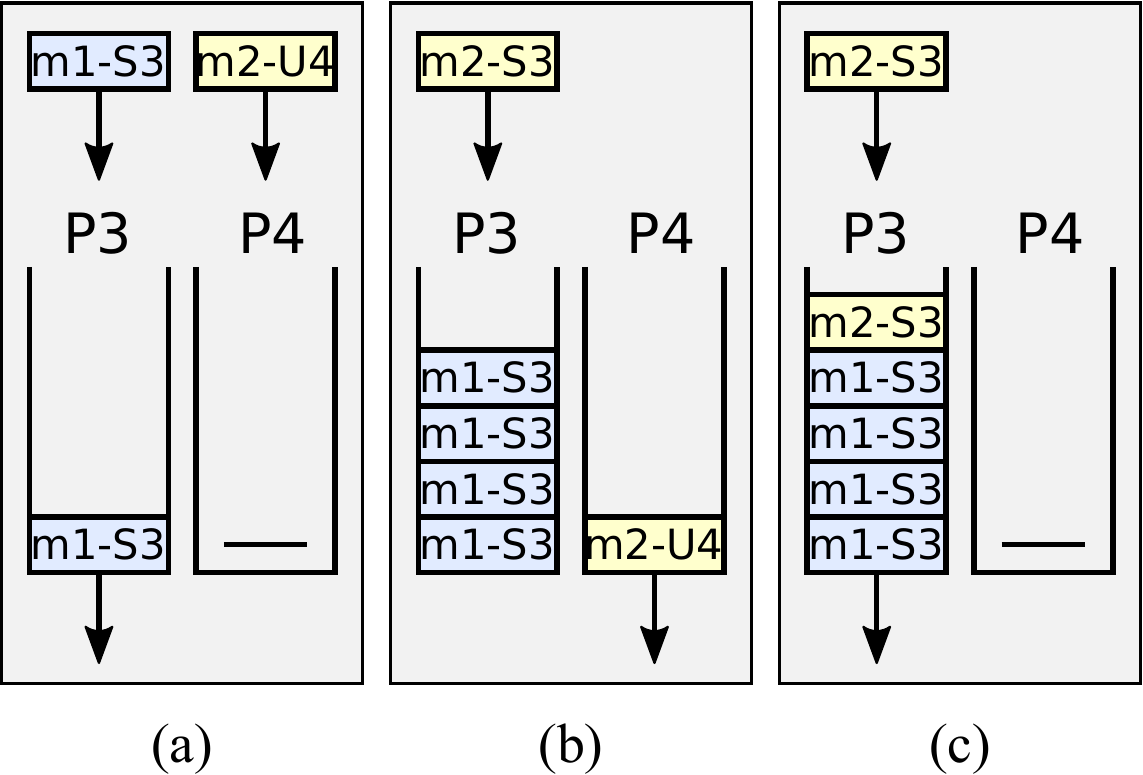}
\caption{Preemption lag occurs if a higher priority message uses
the same priority level as an existing lower priority
message. Packets arrive at the top from the
aggregation switch, pass through the TOR priority queues, and
are transmitted to the receiver at the bottom. The notation
``m1-S3'' refers to a scheduled packet for message \emph{m1} with
priority 3; ``m2-U4'' refers to an unscheduled packet for message
\emph{m2} with priority 4. RTTbytes corresponds to 4 packets.
In (a) the first unscheduled packet for \emph{m2} arrives at the TOR during an
ongoing transmission of scheduled packets for \emph{m1}. Unscheduled
packets have higher priority than scheduled packets, so \emph{m1}'s
scheduled packets will be buffered; (b) shows the state as the
last unscheduled packet for \emph{m2} is being sent to the receiver.
If scheduled packets for \emph{m2} also use priority level 3, they
will be queued behind the buffered packets for \emph{m1} as shown
in (c). If the receiver assigns a higher priority level for \emph{m2}'s
scheduled packets, it avoids preemption lag.
}
\label{fig:preemptLag}
\vspace{\figSqueeze{}}
\end{center}
\end{figure}

For scheduled packets, the receiver specifies a priority in each
GRANT packet, and the sender uses that priority for the granted bytes.
This allows the receiver to dynamically adjust the priority allocation
based on
the precise set of messages being received; this produces a
better approximation to SRPT than approaches such as PIAS, where
priorities are set by senders based on historical trends.
The receiver uses a different priority level for each message,
with higher priorities used for messages with fewer ungranted bytes.
If there are more incoming messages than priority levels,
only the highest priority messages are granted, as described in
\S\ref{sec:overcommit}.
If there are fewer messages than scheduled
priority levels, then Homa uses the lowest of the available priorities;
this leaves higher priority levels free for new higher priority
messages. If Homa always used the highest scheduled priorities, it would
result in \emph{preemption lag}: when a new higher priority message
arrived, its scheduled packets would be delayed by 1 RTT because
of buffered packets from the previous high priority message
(see Figure~\ref{fig:preemptLag}). Using the lowest scheduled
priorities eliminates preemption lag except when all scheduled
priorities are in use.

\subsection{Overcommitment}
\label{sec:overcommit}

One of the important design decisions for Homa is how many incoming
messages a receiver should allow at any given time. A receiver can
stop transmission of a message by withholding grants; once all of the
previously-granted data arrives, the sender will not transmit any
more data for that message until the receiver starts sending grants
again. We use the term \emph{active} to describe the messages for
which the receiver is willing to send grants; the others are
\emph{inactive}.

One possible approach is to keep all incoming messages active at
all times. This is the approach used by TCP and most other
existing protocols. However, this approach results in high buffer occupancy
and round-robin scheduling between messages, both of which contribute
to high tail latency.

In our initial design for Homa, each receiver allowed only one active
message at a time, like pHost. If a receiver had multiple
partially-received incoming messages,
it sent grants only to the highest priority of these; once it had
granted all of the bytes of the highest priority message, it began granting
to the next highest priority message, and so on. The reasoning for
this approach was to minimize buffer occupancy and to implement
run-to-completion rather than round-robin scheduling.

\begin{figure}
\begin{center}
\includegraphics[scale=0.8]{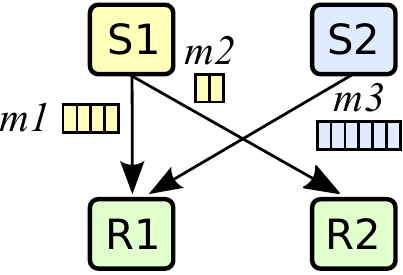}
\caption{Bandwidth can be wasted if a receiver grants to only a single
sender at a time. In this example, S1 has messages ready to send to
R1 and R2 while S2 also has a message for R1. If R1 grants to
only one message at a time, it will choose \emph{m1}, which is shorter
than \emph{m3}.
However, S1 will choose to transmit \emph{m2}, since it is
shorter than \emph{m1}. As a result, R1's downlink will be idle
even though it could be used for \emph{m3}.
}
\label{fig:bipartite}
\vspace{\figSqueeze{}}
\end{center}
\end{figure}

Our simulations showed that allowing only one active message
resulted in poor network utilization under high load.
For example, with workload W4 from Figure~\ref{fig:workloads},
Homa could not use more than about 63\% of the network bandwidth,
regardless of offered load. The network was underutilized because
senders did not always respond immediately to grants; this caused downlink
bandwidth to be wasted. Figure~\ref{fig:bipartite} illustrates how this can
happen.

There is no way for a receiver to know whether a particular sender will
respond to grants, so the only way to keep the downlink fully utilized
is to \emph{overcommit}: a receiver must grant to more
than one sender at a time, even though its downlink can only support
one of the transmissions at a time. With this approach, if one sender
does not respond,
then the downlink can be used for some other sender. If many senders
respond at once, the priority mechanism ensures that the shortest
message is delivered first; packets from the other messages will be
buffered in the TOR.

We use the term \emph{degree of overcommitment} to refer to the
maximum number of messages that may be active at once on a
given receiver.
If there are more than this many messages available, only the
highest priority ones are active. A higher degree of overcommitment
reduces the likelihood of wasted bandwidth, but it consumes more
buffer space in the TOR (up to RTTbytes for each active message)
and it can result in more round-robin scheduling between messages,
which increases average completion time.

Homa currently sets the degree of overcommitment to the number of
scheduled priority levels: a receiver will grant to at most
one message for each available priority level. This approach
resulted in high network utilization in our simulations,
but there are other plausible
approaches. For example, a receiver might use a fixed
degree of overcommitment, independent of available priority levels
(if necessary, several messages could share the lowest priority
level); or, it might adjust the degree of overcommitment dynamically
based on sender response rates.  We leave an exploration of these
alternatives to future work.

The need for overcommitment provides another illustration why it isn't
practical to completely eliminate buffering in a transport
protocol. Homa introduces just enough buffering to ensure good
link utilization; it then uses priorities to make sure that
the buffering doesn't impact latency.

\subsection{Incast}
\label{sec:incast}

Homa solves the incast problem by taking
advantage of the fact that incast is usually self-inflicted: it occurs
when a node issues many concurrent RPCs to other nodes, all of which
return their results at the same time. Homa detects impending
incasts by counting each node's outstanding RPCs. Once this number
exceeds a threshold, new RPCs are marked
with a special flag that causes the server to use a lower
limit for unscheduled bytes in the response message (a few hundred
bytes). Small responses will still get through quickly,
but larger responses will be scheduled by the receiver; the
overcommitment mechanism
will limit buffer usage. With this approach,
a 1000-fold incast will consume at most a few hundred thousand
bytes of buffer space in the TOR.

Incast can also occur in ways that are not predictable;
for example, several machines might simultaneously decide to issue requests
to a single server. However, it is unlikely that many such requests
will synchronize tightly enough to cause incast problems. If this
should occur, Homa's efficient
use of buffer space still allows it to support hundreds of simultaneous
arrivals without packet loss (see Section~\ref{sec:implementationEval}).

Incast is largely a consequence of the high latency in current
datacenters. If each request results
in a disk I/O that takes 10~ms, a client can issue 1000 or more requests
before the first response arrives, resulting in massive incast.
In future low-latency environments, incast will be less of an issue
because requests will complete before very many have been issued.
For example, in the RAMCloud main-memory storage
system~\cite{ramcloud-tocs}, the end-to-end round-trip time for
a read request is about 5$\upmu$s. In a multiread request, it
takes the client 1--2$\upmu$s to issue each request for a different
server; by the time it has issued 3--4 RPCs, responses from the
first requests have begun to arrive. Thus there are rarely more
than a few outstanding requests.

\subsection{Lost packets}

We expect lost packets to be rare in Homa. There are
two reasons for packet loss: corruption in the network, and
buffer overflow. Corruption is extremely rare in modern
datacenter networks, and Homa reduces buffer usage enough
to make buffer overflows extremely uncommon as well. Since packets
are almost never lost, Homa optimizes lost-packet handling for
efficiency in the common case where packets are not lost, and
for simplicity when packets are lost.

In TCP, senders are responsible for detecting lost packets.
This approach requires acknowledgment packets, which add overhead
to the protocol (the simplest RPC requires two data packets and
two acknowledgments). In Homa, lost packets are detected by receivers;
as a result, Homa does not use any explicit acknowledgments. This
eliminates half of the packets for simple RPCs.
Receivers use a simple timeout-based mechanism to detect lost
packets. If a long time period (a few milliseconds) elapses without
additional packets arriving for a message, the receiver sends a RESEND
packet that identifies the first range of missing bytes; the
sender will then retransmit those bytes.

If all of the initial packets of an RPC request are lost, the server
will not know about the message, so it won't issue RESENDs.
However, the client will timeout on the response message, and it
will send a RESEND for the response (it does this even if the request
has not been fully transmitted). When the server receives a
RESEND for a response with an unknown RPCid, it assumes that the request
message must have been lost and it sends a RESEND for the first
RTTbytes of the request.

If a client receives no response to a RESEND (because of server
or network failures), it retries the
RESEND several times and eventually aborts the RPC, returning
an error to higher level software.

\subsection{At-least-once semantics}

RPC protocols have traditionally implemented \emph{at most once}
semantics, where each RPC is executed exactly once in the
normal case; in the event of an error, an RPC may be executed either
once or not at all. Homa allows RPCs to be executed more than once:
in the normal case, an RPC is executed one or more times; after an
error, it could have been executed any number of times (including zero).
There are two situations where Homa re-executes RPCs.
First, Homa doesn't keep connection state,
so if a duplicate request packet arrives after the server has already
processed the original request and discarded its state, Homa will
re-execute the operation. Second, servers get no acknowledgment that
a response was received, so there is no obvious time at which it is
safe to discard the response. Since lost packets are rare, servers
take the simplest approach and discard
all state for an RPC as soon as they have transmitted the last
response packet. If a response packet is lost, the server may
receive the RESEND after it has deleted the RPC state. In this case, it
will behave as if it never received the request and issue
a RESEND for the request; this will result in re-execution of the
RPC.

Homa allows re-executions because it simplifies the implementation
and allows servers to discard all state for inactive clients (at-most-once
semantics requires servers to retain enough state for each client to detect
duplicate requests).
Moreover, duplicate suppression at the transport
level is insufficient for most datacenter applications. For example,
consider a replicated storage system:
if a particular replica crashes while executing a client's request,
the client will retry that request with a different replica.
However, it is possible that the original replica completed the operation
before it crashed. As a result,
the crash recovery mechanism may result in re-execution of a request,
even if the transport implements at-most-once semantics. Duplicates must
be filtered at a level above the transport layer.

Homa assumes that higher level software will either tolerate
redundant executions of RPCs or filter them out. The filtering
can be done either with application-specific mechanisms, or with
general-purpose mechanisms such as RIFL~\cite{rifl}.  For example,
a TCP-like streaming
mechanism can be implemented as a very thin layer on top of
Homa that discards duplicate data and preserves order.

\section{Implementation}
\label{sec:implementation}

We implemented Homa as a new transport in the RAMCloud main-memory
storage system~\cite{ramcloud-tocs}. RAMCloud
supports a variety of transports that use different networking 
technologies, and it has a highly tuned software stack: the total
software overhead to send or receive an RPC is 1--2 $\upmu$s in
most transports. The Homa transport
is based on DPDK~\cite{dpdk}, which allows it to bypass the
kernel and communicate directly with the NIC; Homa detects
incoming packets with polling rather than interrupts. The Homa
implementation contains a total of 3660 lines of C++ code, of 
which about half are comments.

The RAMCloud implementation of Homa includes all of the features described
in this paper except
that it does not yet measure incoming message lengths on the
fly (the priorities were precomputed based on knowledge of the
benchmark workload).

The Homa transport contains one additional mechanism
not previously described, which limits buffer buildup in the NIC
transmit queue. In order for a sender to implement SRPT precisely,
it must keep the transmit queue in the NIC short, so that high
priority packets don't have to wait for lower priority packets
queued previously (as described in \S\ref{sec:senderBehavior},
the sender's priority for an outgoing packet does not necessarily
correspond to the priority stored in the packet). To do this,
Homa keeps a running estimate of the total number of untransmitted
bytes in the NIC, and it only hands off a packet to the NIC
if the number of untransmitted bytes (including the
new packet) will be two full-size packets or less. This allows
the sender to reorder outgoing packets when new messages arrive.

\section{Evaluation}
\label{sec:evaluation}

We evaluated Homa by measuring the RAMCloud implementation and also
by running simulations. Our goal was to answer the following
questions:
\begin{compactitem}
\item Does Homa provide low latency for short messages even at high
network load and in the presence of long messages?
\item How efficiently does Homa use network bandwidth?
\item How does Homa compare to existing state-of-the-art approaches?
\item How important are Homa's novel features to its performance?
\end{compactitem}

\subsection{Implementation Measurements}
\label{sec:implementationEval}

\begin{figure}
\begin{center}
{\footnotesize
\centering
\begin{tabular}{@{}l | p{3cm} | p{3cm}@{}}
& CloudLab & Infiniband \\ \hline
CPU & Xeon D1548 (8 cores @ 2.0~GHz) & Xeon X3470 (4 cores @ 2.93~GHz) \\ \hline
NICs & Mellanox ConnectX-3 (10 Gbps Ethernet) & Mellanox ConnectX-2 (24 Gbps) \\ \hline
Switches & HP Moonshot-45XGc (10 Gbps Ethernet) & Mellanox MSX6036 (4X FDR) and Infiniscale IV (4X QDR) \\
\end{tabular}
}
\vspace{1.0ex}
\caption{Hardware configurations. The Infiniband cluster was used for
measuring Infiniband performance; CloudLab was used for all other
measurements.
}
\label{fig:hardware}
\vspace{\figSqueeze{}}
\end{center}
\end{figure}

We used the CloudLab cluster described in Figure~\ref{fig:hardware}
to measure the performance of the Homa implementation in RAMCloud.
The cluster contained 16 nodes connected to a single switch
using 10 Gbps Ethernet; 8 nodes
were used as clients and 8 as
servers. Each client generated a series of echo RPCs; each RPC sent
a block of a given size to a server, and the server returned the
block back to the client. Clients chose RPC sizes pseudo-randomly
to match one of the workloads from Figure~\ref{fig:workloads}, with
Poisson arrivals configured to generate a particular
network load. The server for each RPC was chosen at random.

\begin{figure*}
\begin{center}
\includegraphics[scale=0.12]{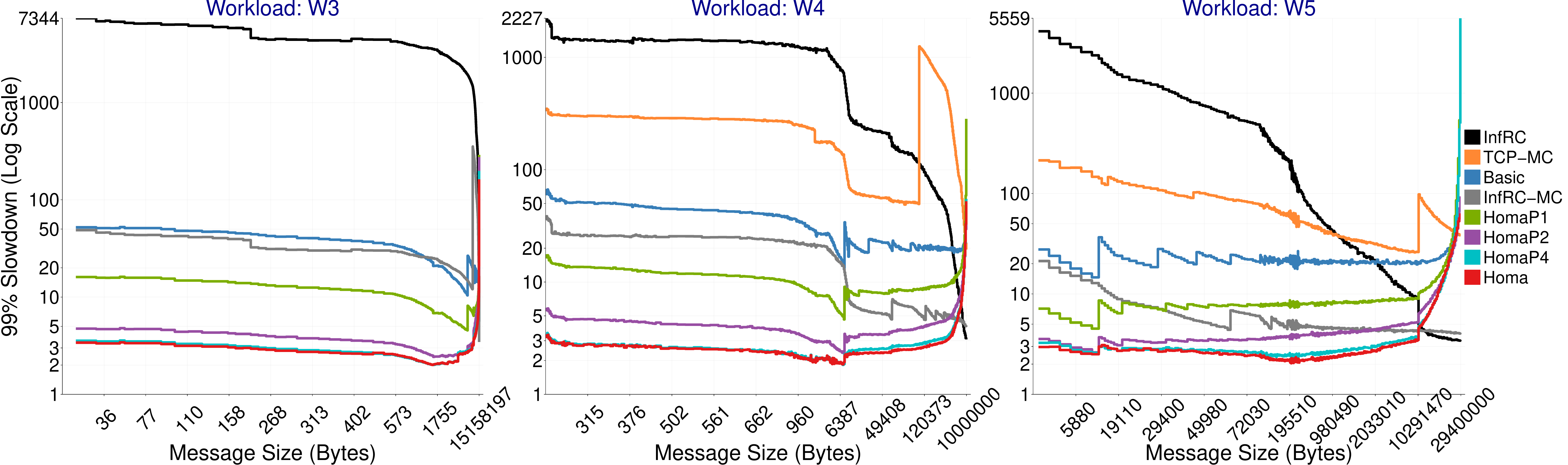}
\caption{Tail latency of Homa and other RAMCloud transports for
workloads W3, W4, and W5 at 80\% network load.
X-axes are linear in total number of messages (each tick is 10\% of
all messages).  ``HomaP\emph{x}'' measures Homa restricted to
use only \emph{x} priorities.
``Basic'' measures
the preexisting Basic transport in RAMCloud, which corresponds
roughly to HomaP1 with no limit on overcommitment. ''InfRC''
measures RAMCloud's Infrc transport, which uses Infiniband reliable
connected queue pairs. ``InfRC-MC'' uses Infiniband with multiple
connections per client-server pair. ``TCP-MC'' uses kernel TCP with
multiple connections per client-server pair. Homa, Basic, and TCP
were measured on the CloudLab
cluster. InfRC was measured on the Infiniband cluster using the
same absolute workload, so its network utilization was only about
33\%. Best-case RPC times (slowdown of 1.0) for 100 byte
RPCs are 3.9~$\upmu$s for InfRC, 4.7~$\upmu$s for Homa and Basic,
and 15.5~$\upmu$s for TCP.
}
\label{fig:slowdownImpl}
\vspace{\figSqueeze{}}
\end{center}
\end{figure*}

\begin{figure*}
\begin{center}
\includegraphics[scale=0.12]{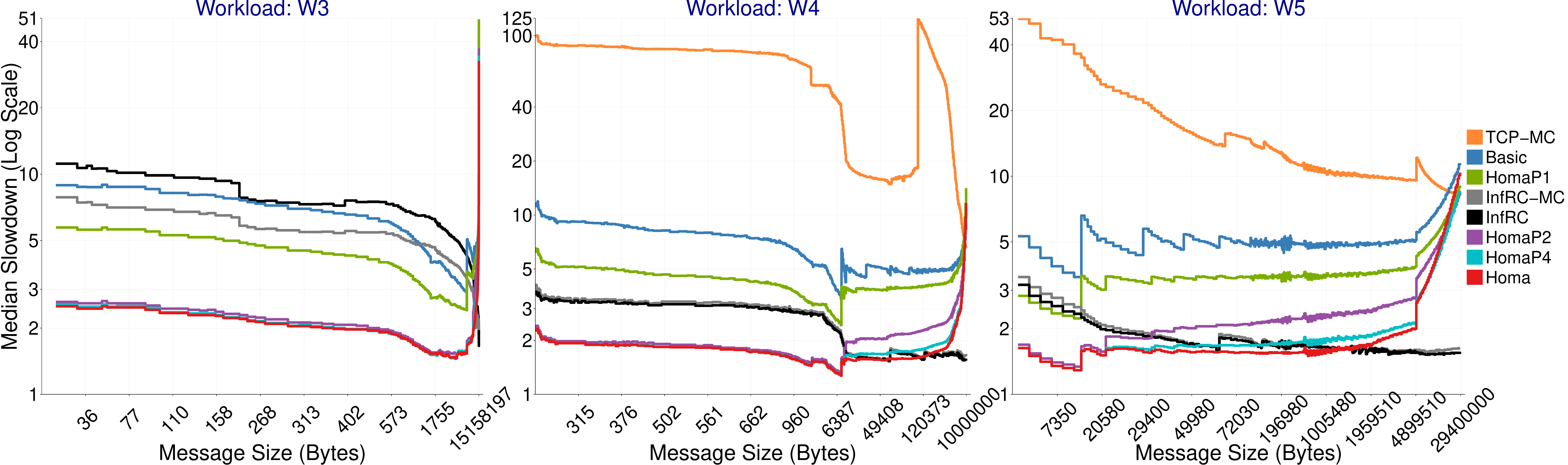}
\caption{Same as Figure~\ref{fig:slowdownImpl} except the y-axis is
median slowdown instead of 99th percentile slowdown.
}
\label{fig:slowdownImpl-median}
\vspace{\figSqueeze{}}
\end{center}
\end{figure*}

Figures~\ref{fig:slowdownImpl} and \ref{fig:slowdownImpl-median} graph the performance of Homa and
several other RAMCloud transports for workloads W3-W5 at 80\%
network load (W1 and W2 are not shown because RAMCloud's software
overheads are too high to handle the large numbers of small messages
generated by these workloads at 80\% network utilization).
Our primary metric for evaluating Homa, shown in
Figure~\ref{fig:slowdownImpl}, is 99th percentile tail
\emph{slowdown}, where slowdown
is the ratio of the actual time required to complete an echo RPC
divided by the best possible time for an RPC of that size on an
unloaded network. A slowdown of 1 is ideal. The x-axis for each
graph is scaled to match the CDF of message sizes: the axis is linear
in total number of messages, with ticks corresponding to 10\% of
all messages in that workload.
This results in a different x-axis scale for each workload,
which makes it easier to see results for the message sizes that
are most common.

Homa provides a 99th percentile tail slowdown in the range of 2--3.5
across a broad range of RPC sizes and workloads. For example, a
100-byte echo RPC takes 4.7 $\upmu$s in an unloaded network; at 80\%
network load, the 99th-percentile latency was about 14 $\upmu$s in
all three loads.

To quantify the benefits provided by the priority and overcommitment
mechanisms in Homa, we also measured RAMCloud's Basic transport.
Basic is similar to Homa in that it is
receiver-driven, with grants and unscheduled packets. However,
Basic does not use priorities and it has no limit on overcommitment:
receivers grant independently to all incoming messages.
Figure~\ref{fig:slowdownImpl} shows that tail latency is 5--15x higher
in Basic than in Homa.
By analyzing detailed packet traces we determined that Basic's high 
latency is caused by queuing delays at the receiver's downlink; Homa's
use of priorities eliminates almost all of these delays.

Although Homa prioritizes small messages, it also outperforms Basic
for large ones. This is because Homa's SRPT policy tends to produce
run-to-completion behavior: it finishes the highest priority message
before giving service to any other messages. In contrast, Basic,
like TCP, tends to produce round-robin behavior; when there are
competing large messages, they all complete slowly.

For the very largest messages, Homa produces 99th-percentile
slowdowns of 100x or more. This is because of the SRPT policy.
We speculate that the performance of these outliers could be
improved by dedicating a small fraction of downlink bandwidth to
the oldest message; we leave a full analysis of this alternative
to future work.

To answer the question ``How many priority levels does Homa need?''
we modified the Homa transport to reduce the number of priority
levels by collapsing adjacent priorities.
Figures~\ref{fig:slowdownImpl} and \ref{fig:slowdownImpl-median}
 show the results. 
99th-percentile tail latency is almost as good with 4 priority levels as
with 8, but tail latency increases noticeably when there
are only 2 priority levels. Even when considering median
slowdown (Figure~\ref{fig:slowdownImpl-median}), performance is
considerably better with two priorities than just one.
Homa with only one priority level is still significantly better
than Basic; this is because Homa's limit on overcommitment
results in less buffering than Basic, which reduces preemption lag.

\smallskip
\noindent{\bf Homa vs. Infiniband.}
Figures~\ref{fig:slowdownImpl} and \ref{fig:slowdownImpl-median} also
measure RAMCloud's InfRC transport,
which uses kernel bypass with Infiniband reliable connected queue pairs.
The Infiniband measurements show the advantage of Homa's message-oriented
protocol over streaming protocols. We first measured InfRC in
its normal mode, which uses a single connection for each client-server
pair. This
resulted in tail latencies about 1000x higher than Homa for small messages.
Detailed traces showed that the long delays were caused by
head-of-line blocking at the sender, where a small message got
stuck behind a very large message to the
same destination. Any streaming protocol, such as TCP, will
suffer similar problems. We then modified the benchmark to use multiple
connections per client-server pair (``InfRC-MC'' in the
figures). This eliminated the head-of-line
blocking and improved tail latency by 100x, to about the same level as Basic.
As discussed in \S\ref{subsec:noConnections}, this approach is probably
not practical in large-scale applications
because it causes an explosion of connection state. InfRC-MC still
doesn't approach Homa's performance, because it doesn't use priorities.

Note: the
Infiniband measurements were taken on a different cluster with faster
CPUs, and the Infiniband network has 24 Gpbs application level
bandwidth, vs. 10 Gbps
for Homa and Basic. The software overheads for InfRC were too high
to run at 80\% load on the Infiniband network, so we used the same
absolute load as for the Homa and Basic measurements, which resulted
in only 33\% network load for Infiniband.
As a result, Figures~\ref{fig:slowdownImpl} and
\ref{fig:slowdownImpl-median} overstate the performance of Infiniband
relative to Homa. In particular, Infiniband appears to perform better
than Homa for large messages sizes. This is an artifact of measuring
Infiniband at 33\% network load and Homa at 80\%; at equal load factors,
we expect Homa to provide significantly lower latency than
Infiniband at all message sizes.

\smallskip
\noindent{\bf Homa vs. TCP.}
The ``TCP-MC'' curves in Figures~\ref{fig:slowdownImpl} and
\ref{fig:slowdownImpl-median} show the
performance of RAMCloud's TCP transport, which uses the Linux
kernel implementation of TCP. Only workloads W4 and W5 are shown
(system overheads were too high to run W3 at 80\% load), and only
with multiple connections per client-server pair (with a single connection,
tail slowdown was off the scale of the graphs).  Even in multi-connection
mode, TCP's tail latencies are 10--100x higher than for
Homa. We also created a new RAMCloud transport using mTCP~\cite{mtcp},
a user-level implementation of TCP that uses DPDK for kernel bypass.
However, we were unable to achieve latencies for mTCP less than 1~ms;
the mTCP developers confirmed that
this behavior is expected (mTCP batches heavily, which improves
throughput at the expense of latency). We did not graph mTCP results.

\smallskip
\noindent{\bf Homa vs. other implementations.}
It is difficult to compare Homa with other published implementations
because most prior systems do not break out small message performance
and some measurements were taken with slower networks. Nonetheless,
Homa's absolute performance (14~$\upmu$s round-trip for small messages
at 80\% network load and 99th percentile tail latency) is nearly
two orders of magnitude faster than the best available comparison
systems. For example, HULL~\cite{hull} reported 782~$\upmu$s one-way
latency for 1 Kbyte messages at 99th percentile and
60\% network load, and PIAS~\cite{pias} reported 2 ms one-way latency for
messages shorter than 100~Kbytes at 99th percentile and 80\% network load;
both of these systems used 1 Gbps networks. NDP~\cite{ndp} reported
more than 600 $\upmu$s one-way latency for 100~Kbyte messages at 99th
percentile in a loaded 10~Gbps network, of which more than 400~$\upmu$s
was queueing delay.

\begin{figure}
\begin{center}
\includegraphics[scale=0.38]{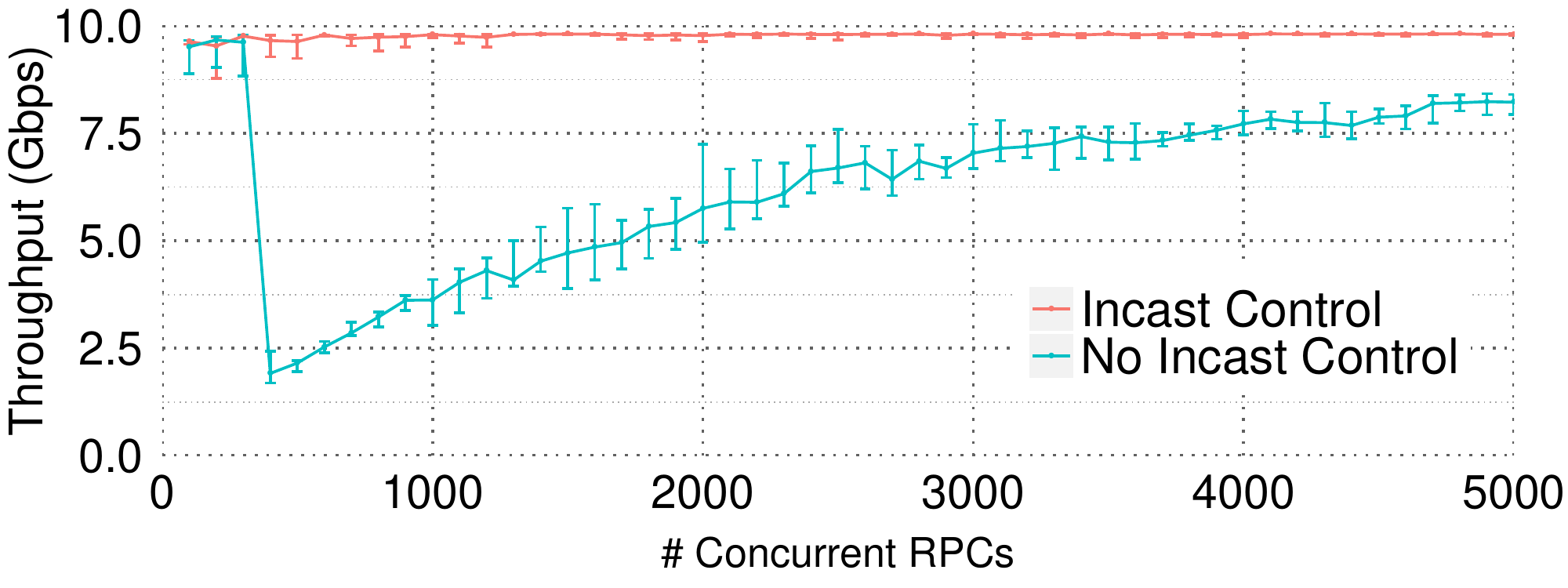}
\caption{Overall throughput when a single Homa client receives responses
for RPCs issued concurrently to 15 servers. Each response was 10 KB. Each data point shows min, mean, and max values over 10 runs.
}
\label{fig:incast}
\vspace{\figSqueeze{}}
\end{center}
\end{figure}

\smallskip
\noindent{\bf Incast.} To measure the effectiveness of Homa's incast
control mechanism, we ran an experiment where a single client initiated
a large number of RPCs in parallel to a collection of servers. Each RPC
had a tiny request and a response of approximately RTTbytes (10 KB).
Figure~\ref{fig:incast} shows the results. With the incast control
mechanism enabled, Homa successfully handled several thousand
simultaneous RPCs without degradation.
We also measured performance with incast control disabled; this
shows the performance that can be expected when incast occurs for
unpredictable reasons. Even under these conditions
Homa supported about 300 concurrent RPCs before performance degraded
because of packet drops. Homa is less sensitive to incast than
protocols such as TCP because its packet scheduling mechanism limits
buffer buildup to at most RTTbytes per incoming message. In contrast,
a single TCP connection can consume all of the buffer space
available in a switch.

\subsection{Simulations}
\label{sec:simulations}

The rest of our evaluation is based on packet-level simulations. The
simulations allowed us to explore more workloads, measure behavior
at a deeper level, and compare with simulations of pFabric~\cite{pfabric},
pHost~\cite{phost}, NDP~\cite{ndp}, and PIAS~\cite{pias}. We chose pFabric
for comparison because it is widely used as a benchmark
and its performance is believed to be near-optimal. We chose pHost and
NDP because they use receiver-driven packet scheduling, like Homa, but
they make limited use of priorities and don't use overcommitment.
We chose PIAS because it uses priorities in a more static fashion than
Homa and does not use receiver-driven scheduling.

\begin{figure}
\begin{center}
\includegraphics[scale=0.7]{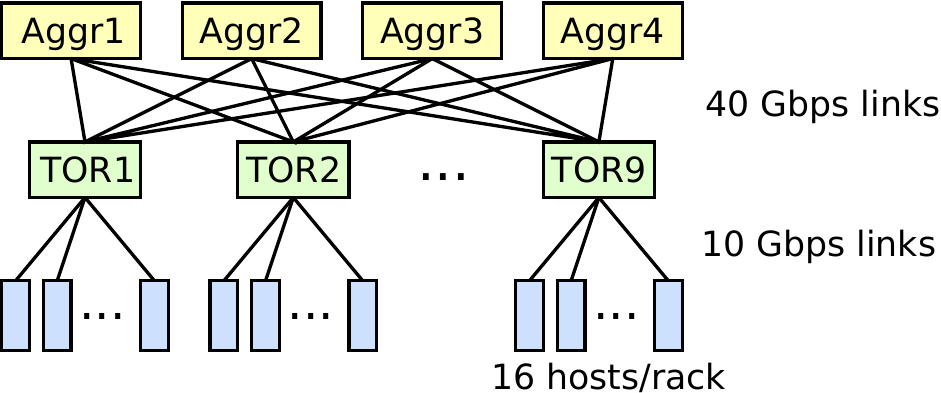}
\caption{The network topology used in simulations. Switches
have an internal delay of 250 ns.
}
\label{fig:topology}
\vspace{\figSqueeze{}}
\end{center}
\end{figure}

We created a packet-level simulator for Homa using the
OMNeT++ simulation framework~\cite{omnet}. We measured pFabric, pHost,
NDP, and PIAS using the simulators developed by their authors. The pFabric
simulator is based on ns-2~\cite{ns2}, and the PIAS simulator is
based on the pFabric simulator. The pHost and NDP simulators were built
from scratch without an underlying framework.
We modified the simulators for pFabric, pHost, NDP, and PIAS to use the
same workloads and network configuration as the Homa simulator.
To the best of our abilities, we tuned each simulator to produce
the best possible performance. The NDP simulator does not support
less-than-full-size packets, so we
used it only for workload W5, in which all packets are full size.

Figure~\ref{fig:topology} shows the network topology used
for simulations, which is the same as that used for prior evaluations
of pFabric, pHost, and PIAS. It consists of 144 hosts
divided among 9 racks with a 2-level switching fabric. Host links
operate at 10 Gbps and TOR-aggregation links operate at 40 Gbps.
The simulated switches do not support cut-through routing.
Speed-of-light propagation delays are assumed to be 0.
The simulations assume that host software has unlimited throughput
(it can process any number of messages per second),
but with a delay of 1.5 $\upmu$s from when a packet arrives at a
host until it has been processed by software and transmission of
a response packet can begin. We chose this delay based on measurements
of the Homa implementation.
The total round-trip time for a receiver to send a small grant packet
and receive the corresponding full-size data packet is thus
about 7.8 $\upmu$s and RTTbytes
is about 9.7 Kbytes (this assumes the two hosts are on different
TORs, so each packet must traverse four links).
The switches implement packet spraying~\cite{packetSpraying}, so
that packets from a given host are distributed randomly across
the uplinks to the core switches.

Our simulations used an all-to-all communication pattern similar
to that of \S\ref{sec:implementationEval}, except that
each host was both a sender and a receiver, and the workload
consisted of one-way messages instead of RPCs. New messages are created at
senders according to a Poisson process; the size of each message is
chosen from one of the workloads in Figure~\ref{fig:workloads}, and
the destination
for the message is chosen uniformly at random. For each simulation
we selected a message arrival rate to produce a desired
network load, which we define
as the percentage of available
network bandwidth consumed by goodput packets; this includes
application-level data plus the minimum overhead (packet headers,
inter-packet gaps,
and control packets) required by the protocol; it does not include
retransmitted packets.

\begin{figure*}
\begin{center}
\begin{minipage}{.5\textwidth}
    \centering
    \begin{center}
    \footnotesize
    \helvetica
    \textbf{80\% Network Load}
    \end{center}
    \includegraphics[scale=0.14]{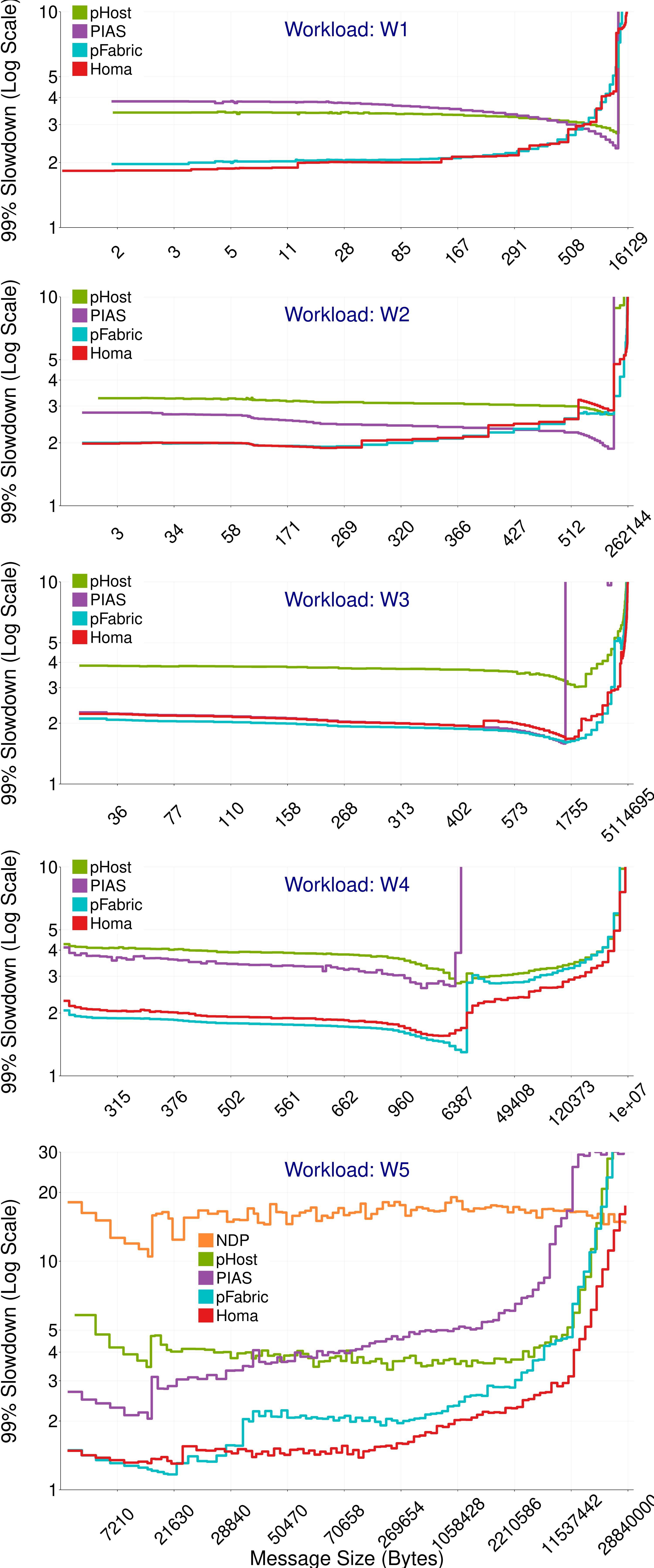}
    \begin{center}
    \footnotesize
    \helvetica
    (a)
    \end{center}
\end{minipage}%
\begin{minipage}{.5\textwidth}
    \centering
    \begin{center}
    \footnotesize
    \helvetica
    \textbf{50\% Network Load}
    \end{center}
    \includegraphics[scale=0.14]{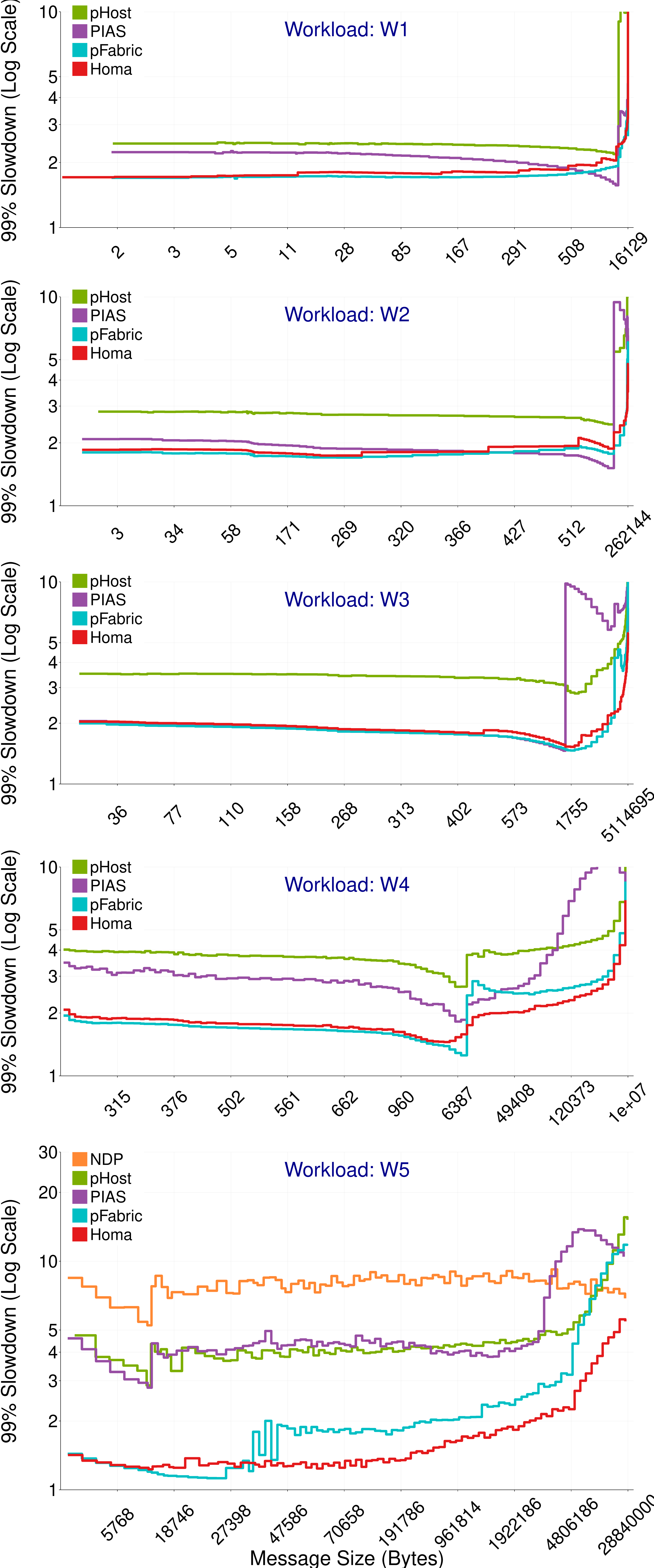}
    \begin{center}
    \footnotesize
    \helvetica
    (b)
    \end{center}
\end{minipage}
\vspace{0.05in}
\caption{99th-percentile slowdown as a function of message size,
for different protocols, workloads, and network loads. Distance on the
x-axis is linear in total number of messages (each tick corresponds
to 10\% of all messages).  Graphs in (a) were measured at
80\% network load, except for NDP and pHost. Neither NDP or pHost
can support 80\% network load for these workloads, so we used the
highest load that each protocol could support (70\% for NDP,
58--73\% for pHost, depending
on workload). Graphs in (b) were measured at 50\% network load.
The minimum one-way time for a small message (slowdown is 1.0) is 2.3 $\upmu$s.
NDP was measured only for W5 because its simulator cannot handle partial
packets.
}
\label{fig:spectrumTail}
\end{center}
\end{figure*}

\begin{figure*}
\begin{center}
\begin{minipage}{.5\textwidth}
    \centering
    \begin{center}
    \footnotesize
    \helvetica
    \textbf{80\% Network Load}
    \end{center}
    \includegraphics[scale=0.14]{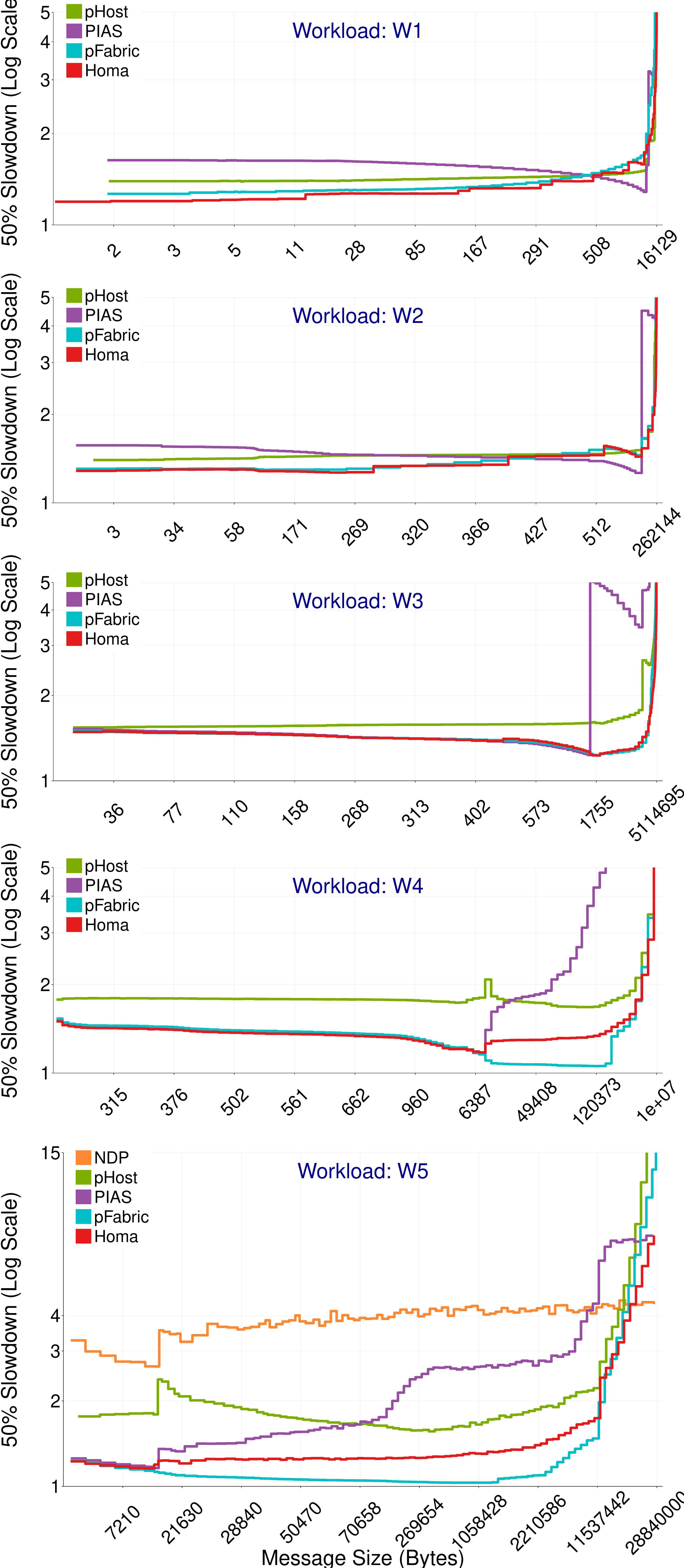}
    \begin{center}
    \footnotesize
    \helvetica
    (a)
    \end{center}
\end{minipage}%
\begin{minipage}{.5\textwidth}
    \centering
    \begin{center}
    \footnotesize
    \helvetica
    \textbf{50\% Network Load}
    \end{center}
    \includegraphics[scale=0.14]{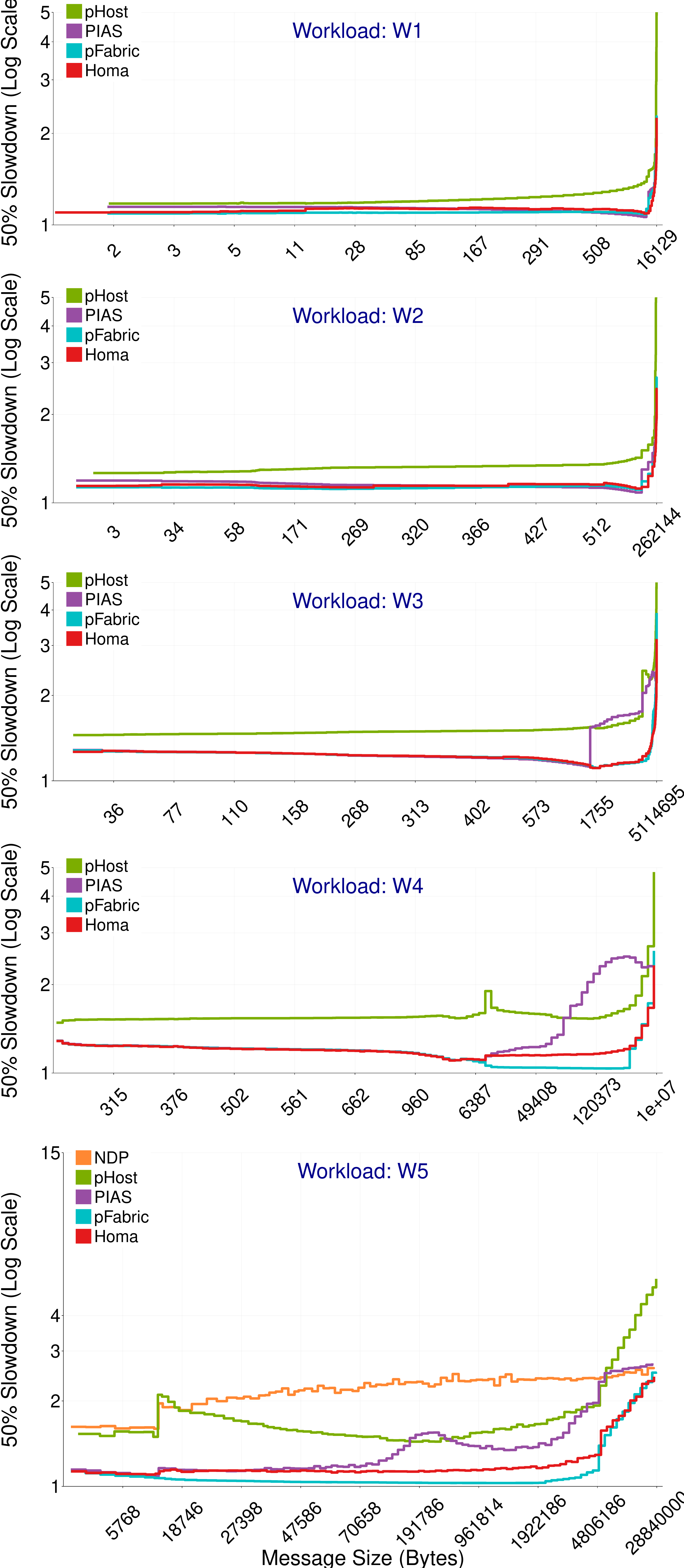}
    \begin{center}
    \footnotesize
    \helvetica
    (b)
    \end{center}
\end{minipage}
\vspace{0.05in}
\caption{Same as Figure~\ref{fig:spectrumTail} except the
y-axis is median slowdown instead of 99th-percentile slowdown.
}
\label{fig:spectrumMedian}
\end{center}
\end{figure*}

\smallskip
\noindent{\bf Tail latency vs. pFabric, pHost, and PIAS.}
Figure~\ref{fig:spectrumTail} displays 99th percentile
slowdown as a function of message size at network loads
of 80\% and 50\%
for the five workloads in Figure~\ref{fig:workloads}.
Figure~\ref{fig:spectrumMedian} displays median slowdown for
the same experiments. Both figures use the same
axes as Figure~\ref{fig:slowdownImpl} except that slowdown
is measured in terms of one-way message delivery, not RPC
round-trips. The discussion below focuses on
Figure~\ref{fig:spectrumTail}(a) (99th percentile at 80\% network
load) because it is the most challenging metric and the one
that motivated Homa's design. The Homa curves in
Figure~\ref{fig:spectrumTail}(a) are similar to those
in Figure~\ref{fig:slowdownImpl}, but slowdowns are somewhat
less in Figure~\ref{fig:spectrumTail}(a) (the simulations do
not model queueing delays that occur in software, such
as when an incoming packet cannot be processed immediately
because the receiver is still processing an earlier packet).

Homa delivers consistent low latency for small messages across
all workloads, and its performance is similar to pFabric:
99th-percentile slowdown for the shortest 50\% of messages is
never worse than 2.2 at 80\% network load. pHost and PIAS have considerably
higher slowdown than Homa and pFabric in
Figure~\ref{fig:spectrumTail}(a). This surprised us, because
both pHost and PIAS claimed performance comparable to pFabric.
On further review, we found that those claims were based on
\emph{mean} slowdown (in Figure~\ref{fig:spectrumMedian} both
pHost and PIAS provide performance closer to pFabric). Our
evaluation follows the original pFabric publication and
focuses on 99th percentile slowdown.

A comparison between the pHost and Homa curves in
Figure~\ref{fig:spectrumTail}(a) shows that a receiver-driven
approach is not enough by itself to guarantee low latency;
using priorities and overcommitment reduces tail latency
by an additional 30--50\%.

The performance of PIAS in Figure~\ref{fig:spectrumTail}(a) is
somewhat erratic. Under most conditions, its tail latency is considerably
worse than Homa, but for larger messages in W1 and W2 PIAS
provides better latency than Homa. PIAS
is nearly identical to Homa for small messages in workload W3.
PIAS always uses the highest priority
level for messages that fit in a single packet, and this happens
to match Homa's priority allocation for W3.

PIAS uses a multi-level feedback queue policy, where each message
starts at high priority; the priority drops as the message
is transmitted and PIAS learns more about its length. This policy
is inferior to Homa's receiver-driven SRPT not only for small messages but
also for most
long ones. Small messages suffer because they get queued behind
the high-priority prefixes of longer messages. Long messages suffer
because their priority drops as they get closer to completion; this makes
it hard to finish them. As a result, PIAS' slowdown jumps significantly
for messages greater than one packet in length. In addition, without
receiver-based scheduling, congestion led to ECN-induced backoff
in workload W4, resulting in slowdowns of 20
or more for multi-packet messages.
Homa uses the opposite approach from PIAS: the priority
of a long message starts off low, but rises as the message gets closer
to finishing; eventually the message runs to completion. In addition,
Homa's rate-limiting and priority mechanisms work well together;
for example, the rate limiter eliminates buffer overflow as a
major consideration.

To show the advantage of SRPT, we made a trivial modification to PIAS.
For short-message workloads such as W1, PIAS allocates
multiple priority levels for the first packet
worth of data. Rather than split the packet, PIAS transmits the
entire packet at the highest priority level. We changed PIAS to
use the \emph{lower} of these priority levels, which makes it
more SRPT-like. With this change, PIAS' performance became nearly
identical to Homa's for messages less than one packet in length.

\smallskip
\noindent{\bf NDP.}
The NDP simulator ~\cite{ndp} could not simulate partial packets,
so we measured NDP only with W5, in which all packets are full-size;
Figure~\ref{fig:spectrumTail}(a) shows the results. NDP's
performance is considerably worse than any of the other protocols, for
two reasons. First, it uses a rate control mechanism with no
overcommitment, which wastes bandwidth: at 70\% network load,
27\% of receiver bandwidth was wasted (the receiver had incomplete
incoming messages yet its downlink was idle). We could not run
simulations above 73\% network load. The wasted downlink bandwidth
results in additional queuing delays at high network load.
Second, NDP does not use SRPT; its receivers use a
fair-share scheduling policy, which results in
a uniformly high slowdown for all messages longer than
RTTbytes. In addition, NDP senders do not prioritize their transmit
queues; this results in severe head-of-line blocking for small messages
when the transmit queue builds up during bursts. The NDP comparison
demonstrates the importance of overcommitment and SRPT.

\begin{figure}
\begin{center}
\vspace{1em}
\includegraphics[scale=0.10]{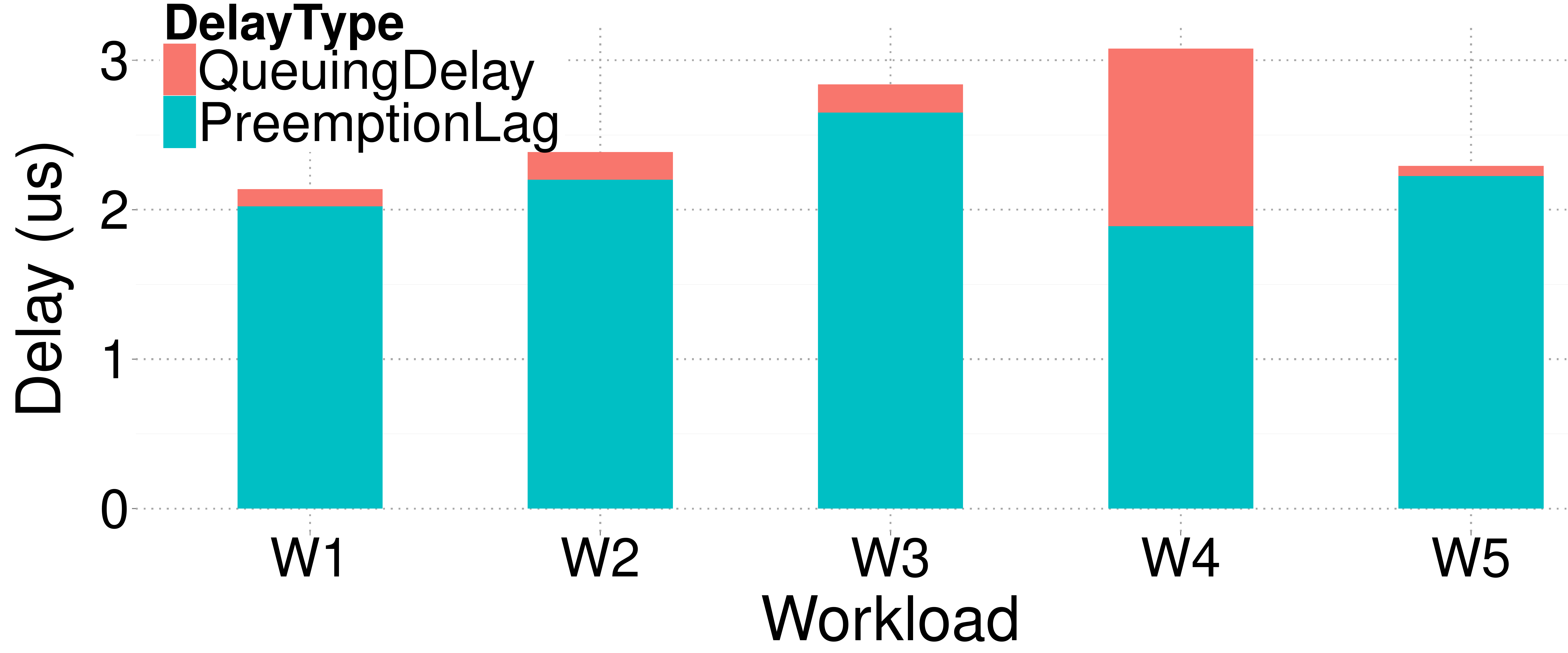}
\caption{Sources of tail delay for short messages. ``Preemption Lag''
occurs when a higher priority packet must wait for a lower priority
packet to finish transmission on a link. ``Queuing Delay" occurs when
a packet waits for one or more packets of equal or higher priority.
Each bar represents an average across short messages with delay
near the 99th percentile. For
workloads W1-W4 the bar considers the smallest 20\% of all messages;
for W5 it considers all single packet messages.
}
\label{fig:tailDelayBrkdown}
\vspace{\figSqueeze{}}
\end{center}
\end{figure}

\smallskip
\noindent{\bf Causes of remaining delay.}
We instrumented the Homa simulator to identify the causes of
tail latency (``why is the slowdown at the 99th percentile
greater than 1.0?'')
Figure~\ref{fig:tailDelayBrkdown} shows that tail latency
is almost entirely due to
link-level preemption lag, where a packet from a
short message arrives at a link while it is busy transmitting a
packet from a longer message. This shows that Homa is nearly optimal:
the only way to improve tail latency significantly is with changes
to the networking hardware, such as implementing link-level
packet preemption.

\begin{figure}
\begin{center}
\includegraphics[scale=0.20]{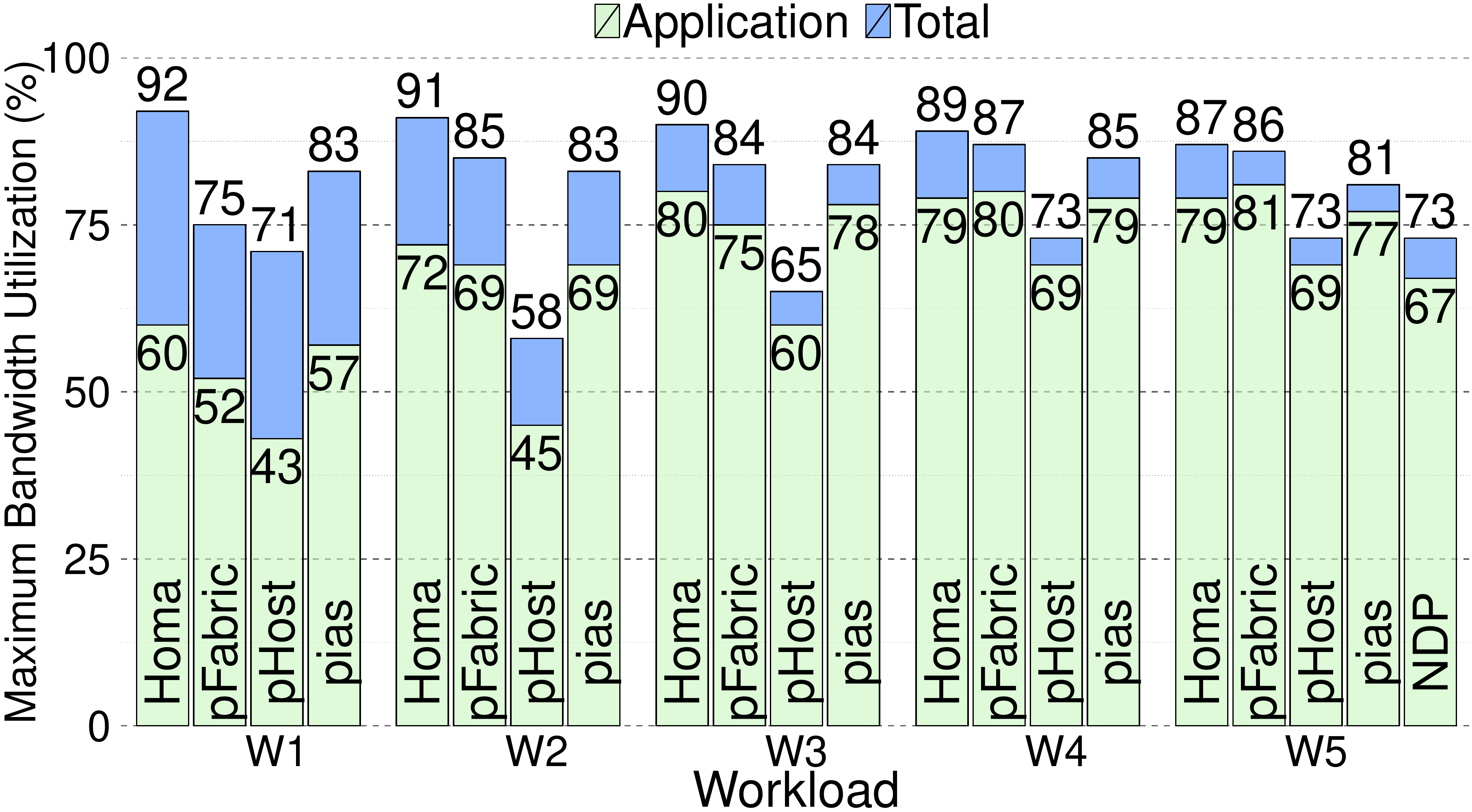}
\caption{Network utilization limits.
The top of each bar indicates the highest percent of available network
bandwidth that the given protocol can support for the given workload.
It counts all bytes in goodput packets, including application data,
packet headers, and control packets; it excludes retransmitted packets.
The bottom part of each bar indicates the percent of network bandwidth
used for application data at that load.
}
\label{fig:bwUtilization}
\vspace{\figSqueeze{}}
\end{center}
\end{figure}

\smallskip
\noindent{\bf Bandwidth utilization.}
To measure each protocol's ability to use network bandwidth efficiently,
we simulated each workload-protocol combination at higher and higher
network loads to identify the highest load the protocol can support
(the load generator
runs open-loop, so if the offered load exceeds the protocol's capacity,
queues grow without bound). Figure~\ref{fig:bwUtilization}
shows that Homa can operate at higher network loads than either
pFabric, pHost, NDP, or PIAS, and its capacity is more stable
across workloads.

\begin{figure}
\begin{center}
\includegraphics[scale=0.38]{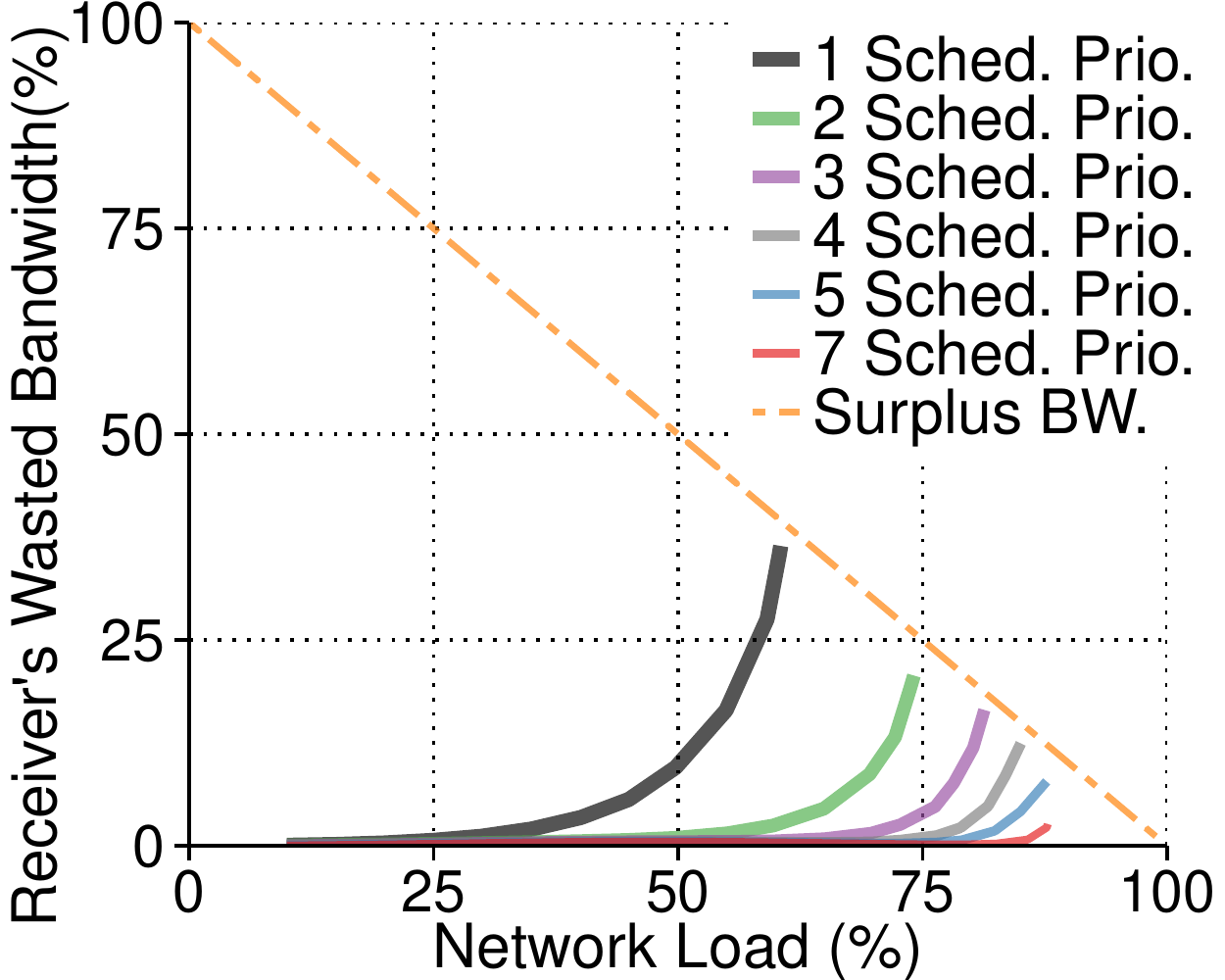}
\caption{Wasted bandwidth as a function of network load for the
W4 workload. Each curve uses a different number of scheduled
priorities, which corresponds to the level of overcommitment.
Each $y$-value is the average fraction of time across all receivers
that a receiver's link is idle, yet the receiver withheld grants
(because of overcommitment limits) that might have caused the
bandwidth to be used. The diagonal line represents
surplus network bandwidth (100\% - network load). Wasted bandwidth
cannot ever exceed surplus bandwidth, so the point where each
curve intersects the diagonal line indicates the maximum sustainable
network load.
}
\label{fig:wastedBw}
\vspace{\figSqueeze{}}
\end{center}
\end{figure}

None of the protocols can achieve 100\% bandwidth because each
of them  wastes network bandwidth under some conditions.
Homa wastes bandwidth because it has a limited number of
scheduled priority levels: there can be times when (a) all of the
scheduled priority levels are allocated, (b) none of those
senders is responding, so the receiver's downlink
is idle and (c) there are additional messages for which the
receiver could send grants if it had more priority levels.
Figure~\ref{fig:wastedBw} shows that this wasted bandwidth
increases with the overall network load; eventually it consumes
all of the surplus network bandwidth. Figure~\ref{fig:wastedBw}
also shows the importance of overcommitment: if receivers grant
to only one message at a time, Homa can only support a network load
of about 63\% for workload W4, versus 89\% with an overcommitment level of 7.

The other protocols also waste bandwidth.
pFabric wastes bandwidth because it drops packets to
signal congestion; those packets must be retransmitted later.
NDP and pHost both waste bandwidth because they do not overcommit
their downlinks. For example, in pHost, if a sender becomes
nonresponsive, bandwidth on
the receiver's downlink is wasted until the receiver times out
and switches to a different sender. Figure~\ref{fig:bwUtilization}
suggests that Homa's overcommitment mechanism uses network bandwidth
more efficiently than any of the other protocols.

\begin{table}
\begin{center}
{
\centering
\small
\begin{tabular}{l l r r r r r}
\textbf{Queue} & & \textbf{W1} & \textbf{W2} & \textbf{W3} & \textbf{W4} & \textbf{W5} \\
\toprule
TOR$\rightarrow$Aggr & mean & 0.7 & 1.0 & 1.6 & 1.7 & 1.7 \\
 & max & 21.1 & 30.0 & 50.3 & 82.7 & 93.6 \\
\noalign{\smallskip}
Aggr$\rightarrow$TOR & mean & 0.8 & 1.1 & 1.8 & 1.7 & 1.6 \\
 & max & 22.4 & 34.1 & 57.1 & 92.2 & 78.1 \\
\noalign{\smallskip}
TOR$\rightarrow$host & mean & 1.7 & 5.5 & 12.8 & 17.3 & 17.3 \\
 & max & 58.7 & 93.0 & 117.9 & 146.1 & 126.4 \\
\noalign{\smallskip}
\end{tabular}
}
\caption{Average and maximum queue lengths (in Kbytes) at switch egress
ports for each of the three levels of the network, measured at 80\%
network load. Queue lengths do not include partially-transmitted or
partially-received packets.
}
\label{tbl:queueLengths}
\end{center}
\end{table}

\smallskip
\noindent{\bf Queue lengths.}
Some queuing of packets in switches is inevitable in Homa
because of its use of unscheduled packets and overcommitment.
Even so, Table~\ref{tbl:queueLengths} shows
that Homa is successful at limiting packet buffering: average queue
lengths at 80\% load are only 1--17 Kbytes, and the
maximum observed queue length was 146 Kbytes (in a TOR$\rightarrow$host
downlink). Of the maximum,
overcommitment accounts for as much as 56 Kbytes (RTTbytes
in each of 6 scheduled priority levels); the remainder is
from collisions of unscheduled packets. Workloads with shorter
messages consume less buffer space than those with longer messages.
For example, the W1 workload uses only one scheduled priority
level, so it cannot overcommit; in addition, its messages are
shorter, so more of them must collide simultaneously in order to
build up long queues at the TOR.
The 146-Kbyte peak occupancy is well within the capacity of
typical switches, so the data confirms our assumption that
packet drops due to buffer overflows will be rare.

Table~\ref{tbl:queueLengths} also validates our assumption that there
will not be significant congestion in the core. The TOR$\rightarrow$Aggr
and Aggr$\rightarrow$TOR queues contain less than 2 Kbytes of data
on average, and their maximum length is less than 100 Kbytes.

\begin{figure}
\begin{center}
\includegraphics[scale=0.12]{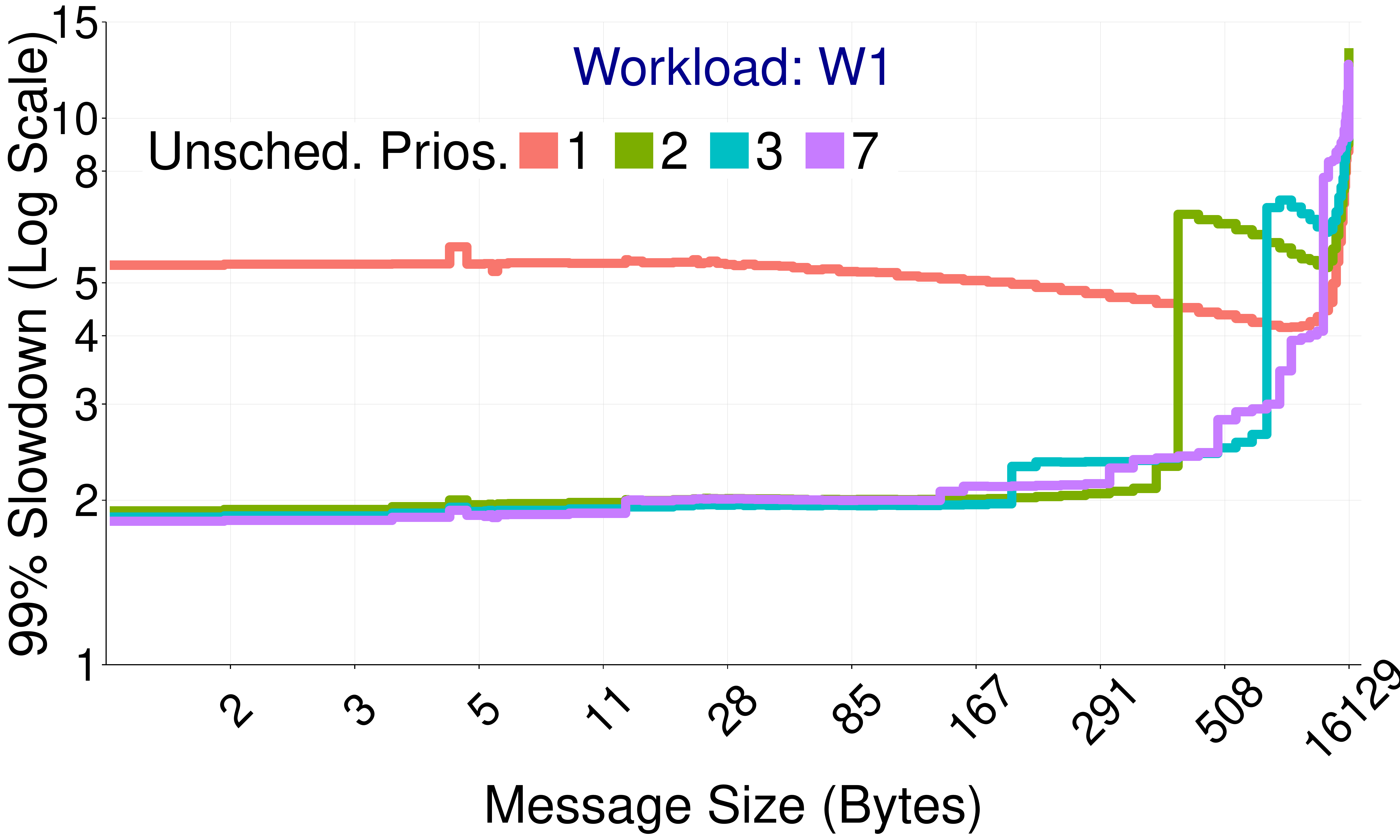}
\caption{Impact of the number of unscheduled priority
levels on workload W1 (80\% network load,
one scheduled priority level). The vertical jumps occur at
the cutoff points between priority levels.
}
\label{fig:slowdownVsUnschedPrios}
\end{center}
\end{figure}

\begin{figure}
\begin{center}
\includegraphics[scale=0.12]{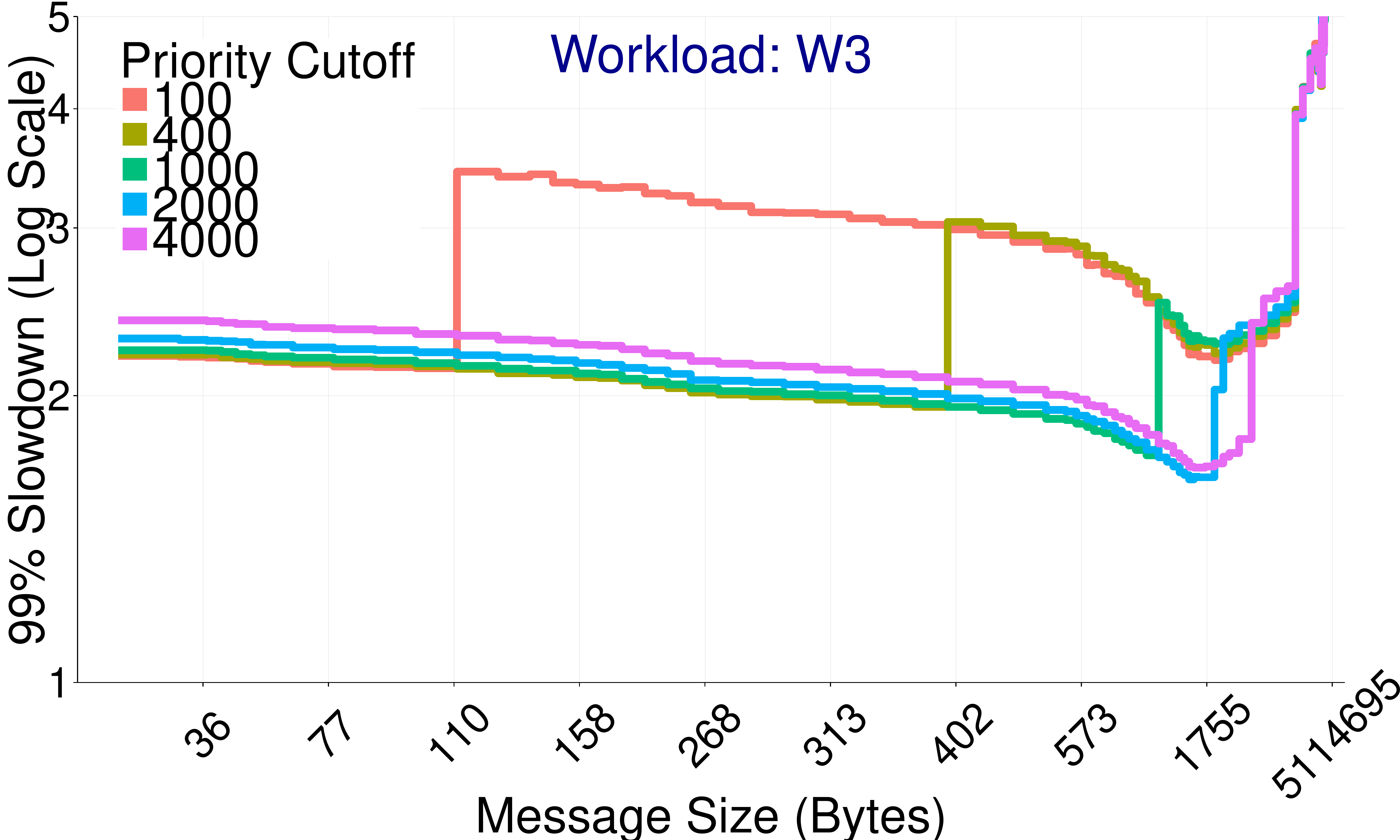}
\caption{The impact of the cutoff point between unscheduled
priorites for workload W3. All measurements were taken at
80\% network load with 2 unscheduled priority levels. Each
curve uses a different cutoff point between the two unscheduled
levels. Homa's algorithm for choosing the cutoff, which balances
the amount of network traffic on the two levels,
would select a cutoff point of 1930 bytes.
}
\label{fig:slowdownVsUnschedPrioCutoff}
\end{center}
\end{figure}

\smallskip
\noindent{\bf Configuration policies.}
Homa automatically configures itself to handle different workloads.
For example,
it allocates 7 priority levels for unscheduled packets in W1,
4 in W3, and only 1 in W4 and W5. In this section we evaluate Homa's
configuration policies by manually varying each parameter in
order to see its impact on performance. For each policy we display
results for the workload with the greatest sensitivity to
the parameter in question.

Figure~\ref{fig:slowdownVsUnschedPrios} shows the
slowdown for workload W1 when the number of unscheduled priorities
was varied from 1 to 7 while fixing the number of scheduled priorities
at 1 (Homa would normally allocate 7 unscheduled priorities for this
workload).
The graph shows that workloads with small messages need multiple
unscheduled priorities
in order to provide low latency: with only a single unscheduled priority,
the 99th percentile slowdown increases by more than 2.5x for most
message sizes. A second priority level improves latency for
more than 80\% of messages; additional priority levels provide
smaller gains.

Figure~\ref{fig:slowdownVsUnschedPrioCutoff} analyzes
Homa's policy for choosing the cutoff points between unscheduled
priority levels. It shows the 99th percentile slowdown for
workload W3 when two priority levels are used for unscheduled
packets and the cutoff point is varied.
Increasing the cutoff point provides a significant latency reduction
for larger messages, while increasing latency slightly for smaller
messages. Up until about 2000 bytes, the penalty for smaller messages
is negligible; however, increasing the cutoff to 4000 bytes results
in a noticeable penalty for about 90\% of all messages, while
providing a large benefit for about 5\% of messages.
Thus, a cutoff of around 2000 bytes provides a reasonable balance.
Homa's policy of balancing traffic in the levels would choose a
cutoff point of 1930 bytes. We considered other
ways of choosing the cutoffs, such
as balancing the number of messages across priority levels; in
Figure~\ref{fig:slowdownVsUnschedPrioCutoff} this would place the
cutoff around 200 bytes, which is clearly sub-optimal.

\begin{figure}
\begin{center}
\includegraphics[scale=0.12]{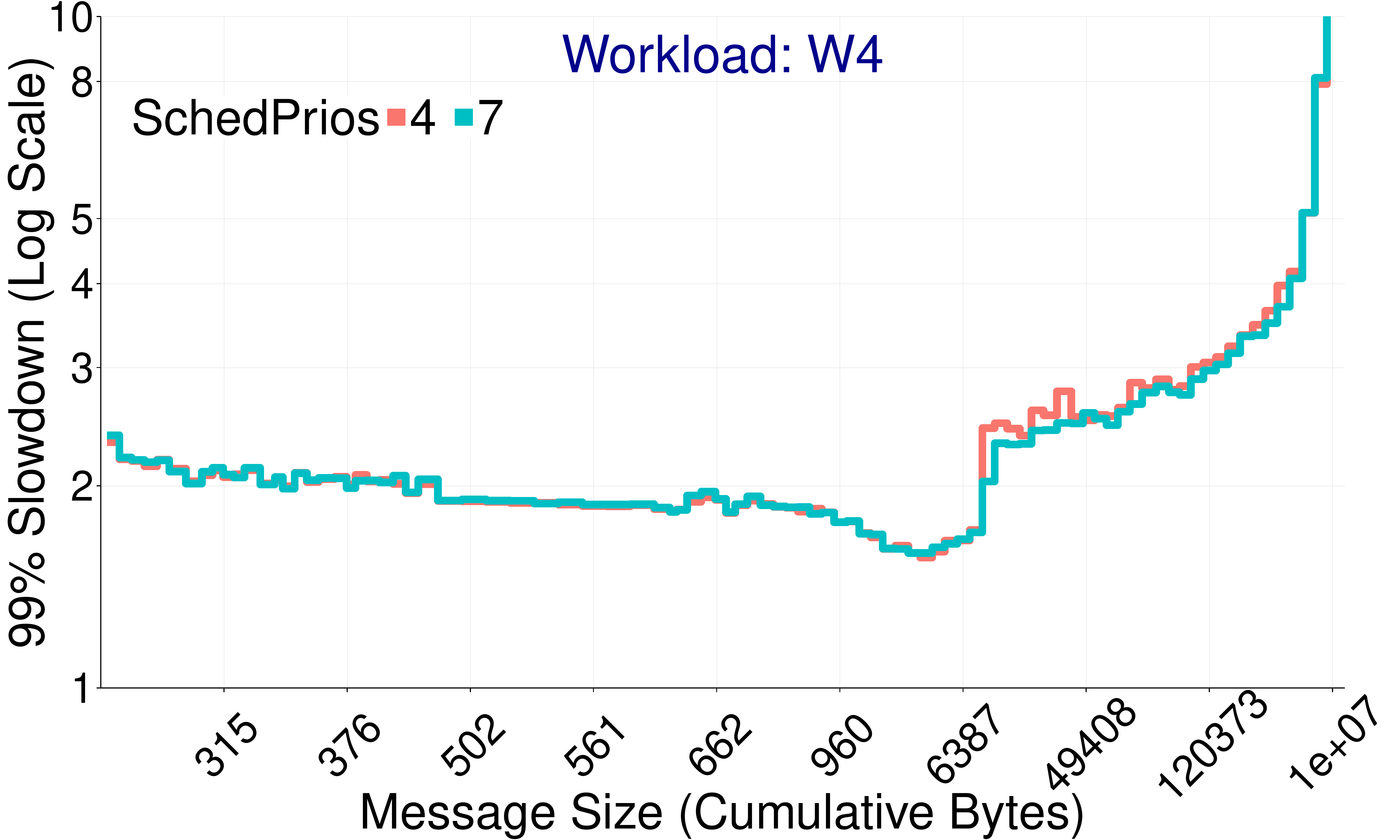}
\caption{Impact of the number of scheduled priority
levels on workload W4 (80\% network load,
one unscheduled priority level).
}
\label{fig:slowdownVsSchedPrios}
\end{center}
\end{figure}

Figure~\ref{fig:slowdownVsSchedPrios} shows the slowdown for
workload W4 with 4 or 7 scheduled priorities, while fixing
the number of unscheduled priorities
at 1 (Homa would normally allocate 7 scheduled priorities for this workload).
Additional scheduled priorities beyond 4 have little impact on
latency. However, the additional scheduled priorities have a significant
impact on the network load that can be sustained (as discussed
earlier for Figure~\ref{fig:wastedBw}). This workload could not run
at 80\% network load with fewer than 4 scheduled priorities.

\begin{figure}
\begin{center}
\includegraphics[scale=0.12]{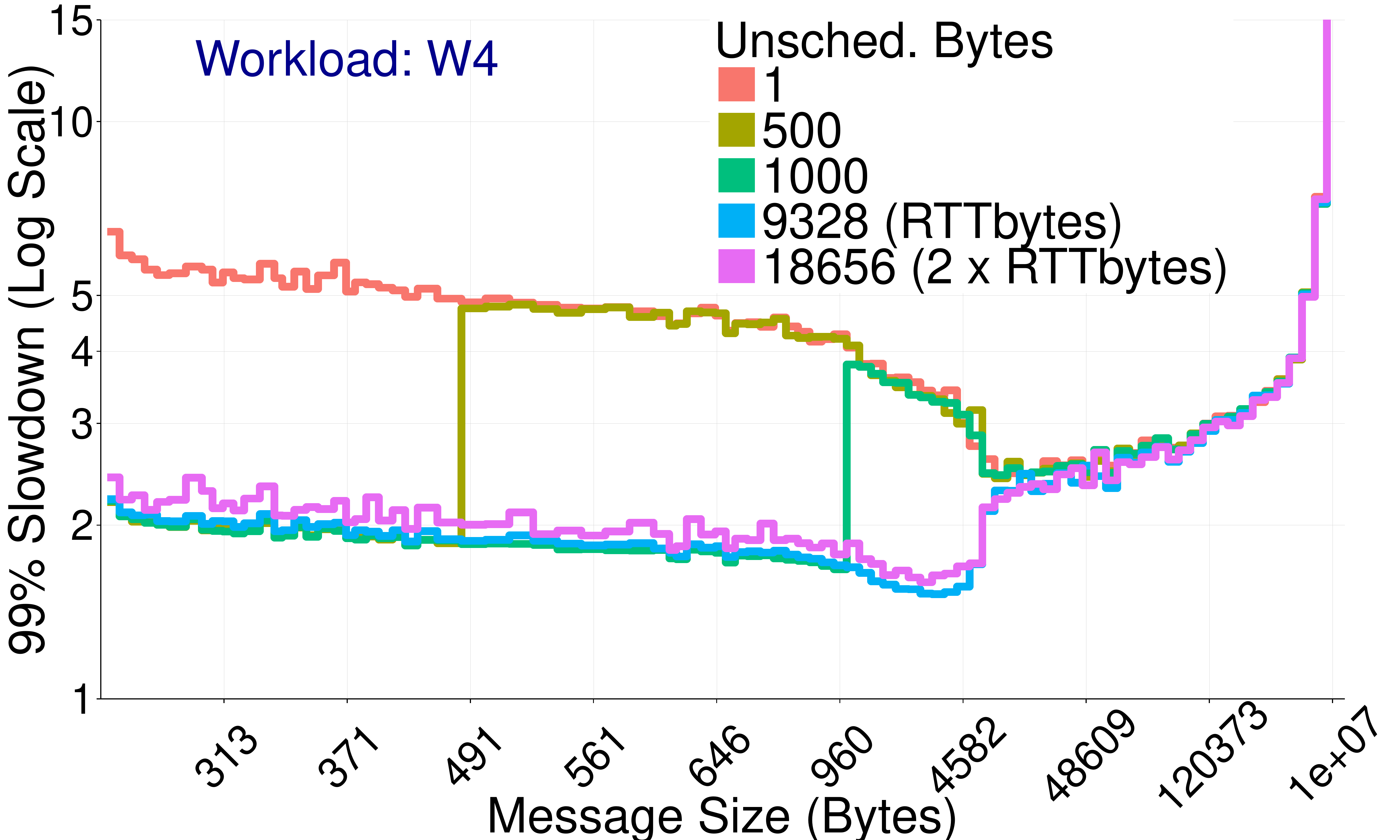}
\caption{The impact of the number of unscheduled bytes on slowdown.
Each curve uses a different limit on the number of unscheduled bytes per message.
All measurements used workload W4 at 80\% network load.
}
\label{fig:slowdownVsUnsched}
\end{center}
\end{figure}

Figure~\ref{fig:slowdownVsUnsched} shows the slowdown for workload
W4 when the number of unscheduled bytes per message is varied.
The figure demonstrates the benefits of unscheduled packets:
messages smaller than RTTbytes but larger than the unscheduled
limit suffer 2.5x worse latency. Increasing the
unscheduled limit beyond RTTbytes results in worse performance
for messages smaller than RTTbytes, because of additional
traffic sharing the single unscheduled priority level.

\begin{figure}
\begin{center}
\vspace{1em}
\includegraphics[scale=0.45]{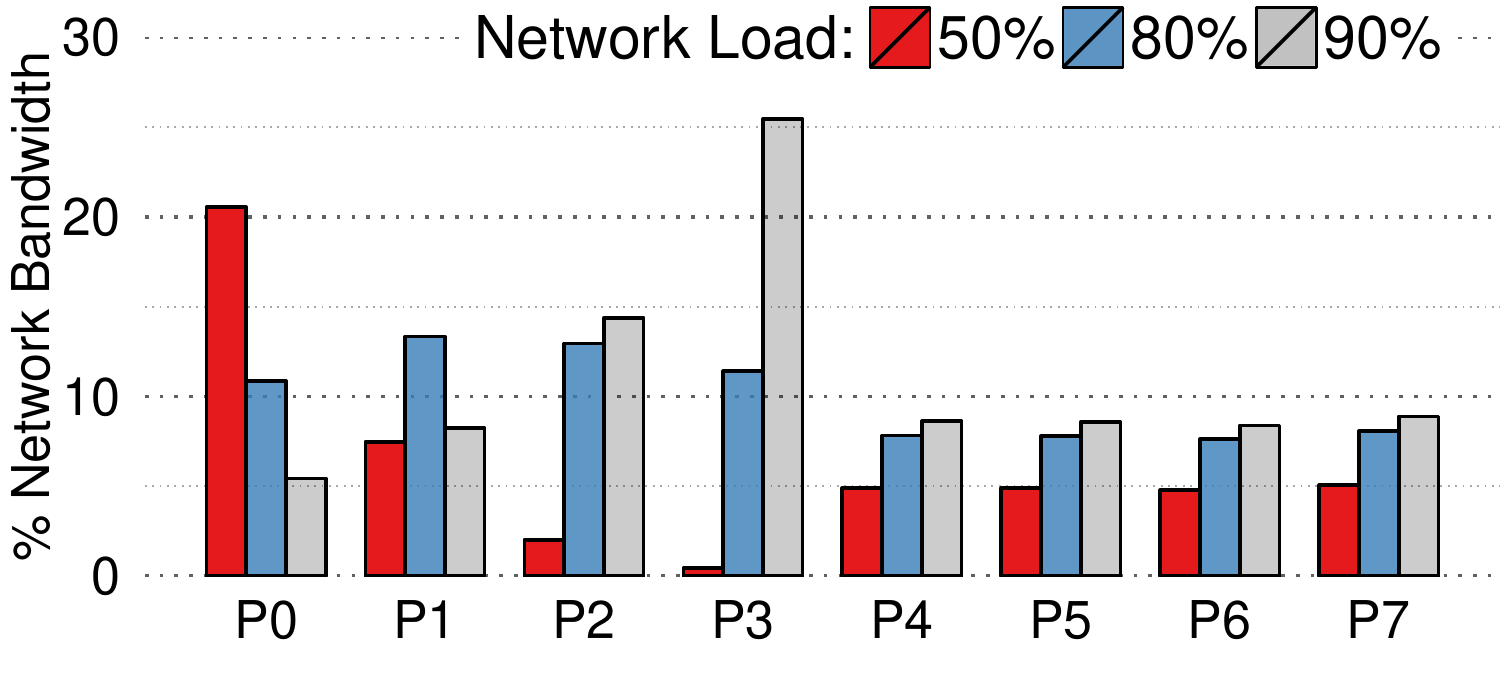}
\caption{Usage of priority levels for workload W3 under different
loads. Each bar indicates the number of network bytes transmitted at
a given priority level, as a fraction of total available network
bandwidth. P0-P3 are used for scheduled packets and P4-P7 for
unscheduled packets.
}
\label{fig:prioUsage}
\end{center}
\end{figure}

\smallskip
\noindent{\bf Priority utilization.}
Figure~\ref{fig:prioUsage} shows how network traffic is divided
among the priority
levels when executing workload W3 at three different network
loads. For this workload Homa splits the priorities evenly between
scheduled and unscheduled packets. The four unscheduled priorities
are used evenly, with the same number of network bytes transmitted
under each priority level. As the network load increases, the
additional traffic is also split evenly across the unscheduled
priority levels.

The four scheduled priorities are used in different ways depending
on the network load. At 50\% load, a receiver typically has only one
schedulable message at a time, in which case the message uses
the lowest priority level (P0). Higher priority levels are used for preemption
when a shorter message appears part-way through the reception of a
longer one. It is rare for preemptions to nest deeply enough to use
all four scheduled levels. As the network load increases, the usage
of scheduled priorities changes. By the time network load
reaches 90\%, receivers
typically have at least four partially-received messages at any given time,
so they use all of the scheduled priority levels. More scheduled packets
arrive on the highest scheduled level than any other; the other levels
are used if the highest-priority sender is nonresponsive or
if the number of incoming messages drops below 4. The figure indicates
that senders are frequently nonresponsive at 80\% network load (more
than half of the scheduled traffic arrives on P0--P2).

\section{Limitations}
\label{sec:limitations}

This section summarizes the most important assumptions Homa makes
about its operating
environment. If these assumptions are not met, then Homa may not achieve
the performance levels reported here.

Homa is designed for use in datacenter networks and capitalizes
on the properties of those networks; it is unlikely to work well in
wide-area networks.

Homa assumes that congestion occurs
primarily at host downlinks, not in the core of the network.
Homa assumes per-packet spraying to ensure load balancing across core
links, combined with sufficient overall capacity. Oversubscription
is still possible, as long as there is enough aggregate bandwidth
to avoid significant congestion. We hypothesize that congestion in
the core of datacenter networks will be uncommon because it will not be
cost-effective. If the core is congested, it will result in
underutilization of servers, and the cost of this underutilization will
likely exceed the cost of provisioning more core bandwidth. If the core does
become congested, then Homa latencies will degrade. Homa's mechanisms
for limiting buffer occupancy may reduce the impact of congestion
in comparison to TCP-like protocols, but we leave a full exploration of
this topic to future work.

Homa also assumes a single implementation of the protocol for each
host-TOR link, such as in an operating system kernel running on bare hardware,
so that Homa is aware of all incoming and outgoing traffic.  If multiple
independent Homa implementations share a single host-TOR link,
they may make conflicting decisions. For example, each
Homa implementation will independently overcommit the downlink and
assign priorities based on the input traffic passing through that
implementation. Multiple implementations can occur when a virtualized
NIC is shared between multiple guest operating systems in a virtual
machine environment, or between multiple applications that implement
the protocol at user level. Obtaining good performance in these
environments may require sharing state between the Homa implementations,
perhaps by moving part of the protocol to the NIC or even the TOR.
We leave an exploration of this problem and its potential solutions to
future work.

Homa assumes that the most severe forms of incast are predictable
because they are self-inflicted by outgoing RPCs; Homa handles these
situations effectively. Unpredictable incasts can also occur, but Homa
assumes that they are unlikely to have high degree. Homa can
handle unpredictable incasts of several hundred messages with typical
switch buffer capacities; unpredictable incasts larger than this will
cause packet loss and degraded performance.

The Homa configuration and measurements in this paper were based on
10 Gbps link speeds. As link speeds increase in the future, RTTbytes
will increase proportionally, and this will impact the protocol in
several ways. A larger fraction of traffic will be sent unscheduled, so
Homa's use of multiple priority levels for unscheduled
packets will become more important.  With faster networks, workloads will
behave more like W1 and W2 in our measurements,
rather than W3-W5. As RTTbytes increases, each message can
potentially consume more
space in switch buffers, and the degree of unpredictable incast that
Homa can support will drop.
\section{Related Work}
\label{sec:related}

In recent years there have been numerous proposals for new transport
protocols, driven by new datacenter applications and the well-documented
shortcomings of TCP. However, none of these proposals combines
the right set of features to produce low latency for short messages under
load.

The biggest shortcoming of most recent proposals is that they do not
take advantage of in-network priority queues. This includes rate-control
techniques such as DCTCP~\cite{dctcp} and HULL~\cite{hull}, which reduce
queue occupancy, and D\textsuperscript{3}~\cite{d3} and
D\textsuperscript{2}TCP~\cite{d2tcp}, which incorporate deadline-awareness.
PDQ~\cite{pdq} adjusts flow rates to implement preemption, but its
rate calculation is too slow for scheduling short
messages. Without the use of priorities, none of these systems can
achieve the rapid preemption needed by short messages.

A few systems have used in-network priorities, but they do not
implement SRPT. \S\ref{sec:simulations} showed that the PIAS
priority mechanism \cite{pias} performs worse than SRPT
for most message sizes and workloads.
QJUMP~\cite{qjump} requires priorities to be
specified manually on a per-application basis. Karuna~\cite{karuna}
uses priorities to separate deadline and non-deadline flows, and
requires a global calculation for the non-deadline flows.
Without receiver-driven SRPT, none of these systems can achieve
low latency for short messages.

pFabric~\cite{pfabric} implements SRPT by assuming fine-grained priority
queues in network switches. Although this produces near-optimal
latencies, it depends on features not available in existing
switches.

pHost~\cite{phost} and NDP~\cite{ndp} are the systems most similar to
Homa, in that both use receiver-driven scheduling and priorities.
pHost and NDP use only two priority levels with
static assignment, which results in poor latency for short messages.
Neither system uses overcommitment, which limits their ability to operate
at high network load. NDP uses fair-share scheduling rather than SRPT,
which results in high tail latencies.
NDP includes an
incast control mechanism, in which network switches drop all but the
first few bytes of incoming packets when there is congestion.
Homa's incast control mechanism achieves a similar effect using a
software approach: instead of truncating packets in-flight
(which wastes network bandwidth),
senders are instructed by the protocol to limit how much data
they send.

Almost all of the systems mentioned above, including DCTCP, pFabric,
PIAS, and NDP, use a connection-oriented streaming approach. As previously
discussed, this results in
either high tail latency because of head-of-line blocking at senders, or an
explosion of connections, which is impractical for large-scale datacenter
applications.

A final alternative is to schedule all messages or packets for a
cluster centrally, as in Fastpass~\cite{fastpass}. However, communication
with the central scheduler adds too much latency to provide good
performance for short messages.
In addition, scaling a system like Fastpass to a large cluster
is challenging, particularly for workloads with many short messages.

\section{Conclusion}
\label{sec:conclusion}

The combination of tiny messages and low-latency networks creates
challenges and opportunities that have not been addressed by
previous transport protocols. Homa meets this need with a new transport
architecture that combines several unusual features:
\begin{compactitem}
\item It implements discrete messages for remote procedure calls, not
byte streams.
\item It uses in-network priority queues with a hybrid allocation
mechanism that approximates SRPT.
\item It manages most of the protocol from the receiver, not the sender.
\item It overcommits receiver downlinks in order to maximize throughput at
high network loads.
\item It is connectionless and has no explicit acknowledgments.
\end{compactitem}
These features combine to produce nearly optimal latency for short messages
across a variety of workloads.
Even under high loads, tail latencies are within a small factor of
the hardware limit.
The remaining delays are almost entirely due to the absence
of link-level packet preemption in current networks; there is
little room for improvement in the protocol itself.
Finally, Homa can be implemented with no changes to networking hardware.
We believe that Homa provides
an attractive platform on which to build low-latency datacenter
applications.
\section{Acknowledgments}
\label{sec:acks}

This work was supported by C-FAR (one of six centers of
STARnet, a Semiconductor Research Corporation program,
sponsored by MARCO and DARPA) and by the industrial affiliates
of the Stanford Platform Laboratory.
Amy Ousterhout, Henry Qin, Jacqueline Speiser, and 14 anonymous
reviewers provided helpful comments on drafts of this paper.
We also thank our SIGCOMM shepherd, Brighten Godfrey.

\bibliographystyle{abbrv}
\bibliography{local}

\end{document}